\documentclass[12pt]{article}
\usepackage{psfrag}
\usepackage{epsfig}
\usepackage{graphicx}
\usepackage{epsf}
\usepackage{feynmp}
\input epsf.tex

\newcommand{\del}{\partial}
\newcommand{\tr}{\mbox{tr}}
\newcommand{\MB}[3]{{\cal M}\left( #1, #2, #3 \right)}
\newcommand\eps{\epsilon}

\newcommand\be{\begin{equation}}
\newcommand\ee{\end{equation}}
\newcommand\barr{\begin{array}}
\newcommand\earr{\end{array}}
\newcommand\ba{\begin{eqnarray}}
\newcommand\ea{\end{eqnarray}}

\def\AP#1{Annals of Physics~{\bf #1}}

\def\IJMP#1{Int. J. Mod. Phys.~{\bf #1}}

\def\NP#1{Nucl. Phys.~{\bf #1}}

\def\PH#1{Physica {\bf #1}}
\def\PL#1{Phys. Lett.~{\bf #1}}
\def\PR#1{Phys. Rev.~{\bf #1}}
\def\PRP#1{Phys. Rep.~{\bf #1}}
\def\PRL#1{Phys. Rev. Lett.~{\bf #1}}

\def\RMP#1{Rev. Mod. Phys.~{\bf #1}}

\def\ZP#1{Z. Phys.~{\bf #1}}

\def\mpla#1{Mod. Phys. Lett. {\bf A#1}}

\def\zpc#1{Z. Phys. {\bf C#1}}
 
\def\be{\begin{eqnarray}}
\def\ee{\end{eqnarray}}

\def\del{\partial}

\newcommand{\f}[2]{\frac{#1}{#2}}

\begin{fmffile}{grap}
\begin{document}

\title{\bf Renormalization Group Flow Equations and the Phase Transition in $O(N)$--models} 
 
\author{
O. Bohr,
B.-J. Schaefer\\
and\\
J. Wambach\\[2mm]
 {\small \em {Institut f\"{u}r Kernphysik, TU Darmstadt, 
        D-64289 Darmstadt, Germany}}
}

\maketitle
\begin{abstract}
\noindent
We derive and solve flow equations for a general $O(N)$-symmetric
effective potential including wavefunction renormalization corrections
combined with a heat-kernel regularization. We investigate the model
at finite temperature and study the nature of the phase transition in
detail. Beta functions, fixed points and critical exponents $\beta$,
$\nu$, $\delta$ and $\eta$ for various $N$ are independently
calculated which allow for a verification of universal scaling
relations.
\end{abstract}

\section{Introduction}
\label{introduction}

The manifestation of universality of critical phenomena enables the
applicability of the massless $O(N)$ scalar field theories to a wide
class of very different physical systems in the vicinity of the
critical temperature $T_c$. E.g.~for $N=4$ this theory is used as an
effective model (the linear $\sigma$-model) for the chiral phase
transition in two light-quark flavor QCD which may have the same
universality class \cite{meye}.  On the other hand in condensed matter
physics the case $N=3$ corresponds to the well-known Heisenberg model
describing the ferromagnetic phase transition. Further applications
(like the Kosterlitz-Thouless phase transition ($N=2$), liquid-vapor
transition describable by the continuous spin Ising model ($N=1$) or
even the statistical properties of long polymer chains ($N=0$)) are
often discussed in the literature.

In the vicinity of a second-order phase transition the equation of
state (EoS) obeys a universal scaling form, which cannot be assessed
by ordinary perturbation theory due to infrared divergences. For a
negative mass term the $N$-component $\Phi^4$ theory exhibits
spontaneous symmetry breaking of the $O(N)$-group down to $O(N-1)$
which is restored at sufficiently high temperature.  According to the
Goldstone-theorem ($N-1$) fields (the pions) become massless in the
chiral limit and the remaining one (the radial sigma mode) stays
massive. Due to these massless modes a naive perturbation expansion
for the effective potential fails to give an appropriate description
and the potential itself develops an imaginary part around the
origin. By means of a nonperturbative Wilsonian renormalization group
approach one can circumvent these difficulties and go beyond any
finite order in perturbation theory.

In this work we present a renormalization group (RG) approach
formulated in the Euclidean space. It is combined with a heat-kernel
regularization which allows a straightforward calculation of the
effective potential with an $O(N)$-symmetry for arbitrary $N$, not
only at finite temperature but also directly at the critical
temperature. We determine several critical exponents and investigate
the phase transition regime in detail going beyond the so-called
'local potential approximation' to also include wavefunction
renormalization corrections.

The paper is organized as follows: in Sect.~2 we summarize the
Wilsonian renormalization group concept and show how flow equations
combined with a proper-time (heat-kernel) regularization are obtained
for a general $O(N)$-symmetric potential. Based on a perturbative
one-loop effective potential expression we derive flow equations for
the potential and wavefunction renormalization for a $d$-dimensional
massless $O(N)$ scalar theory. These coupled RG flow equations are
further improved by taking into account the continuous feedbacks from
the higher modes to the lower ones. The derivation of the flow
equations with wavefunction renormalization corrections is given in
Sect.~\ref{section3}.

Sect.~\ref{section4} details the numerical results of the solution of
the flow equations. We start with a discussion of the numerical
implementation of the properly rescaled equations for a general
potential with $O(N)$-symmetry on a grid whereby the full potential 
is numerically taken into account without resorting to polynomial 
approximations.  The order parameter and the potential itself are 
studied at finite temperature.

Sect.~\ref{section5} is dedicated to the critical behavior of the
system at the transition temperature. The scaling solution, fixed
points and beta functions are investigated in detail and numerical
calculations of the critical exponents for different $N$ including the
anomalous dimensions are presented. They are found to be in perfect
agreement with other works and approaches. Sect.~\ref{conclusion}
contains the summary and concludes our discussion.  Technical details
are relegated to the appendix in order to improve the readability.

\section{Concept of the RG method}
\label{floweqs}

The effective action $\Gamma [\Phi]$ is defined by a Legendre
transformation of the generating functional $W[J]$, which generates
all connected Feynman diagrams and can be expressed by a functional
path integral. Except for numerical simulations, these integrals
usually cannot be calculated exactly.  In addition they are plagued by
infrared (IR) divergences. Instead of performing the integration in
one step one can follow the idea by Wilson and Kadanoff \cite{wils}
and split the field $\Phi$ into fast-fluctuating and slowly-varying
modes
\be
\Phi (x) = \left.\Phi_{\mbox{\small slow}} (x)\right|_{q\le k} +
\left. \Phi_{\mbox{\small fast}} (x)\right|_{q\ge k}
\ee
and perform the functional integration only over the fast-fluctuating
components with momenta $q$ larger than the scale $k$ separating both
field modes from each other.  The irrelevant short-distance modes
which are insensitive to the physics at large correlation scales are
thus eliminated.  One is left with a low-energy ``effective action''
parameterized by the averaged fields at the IR scale $k$ which acts as
a momentum cutoff. The full effective action describing the
one-particle-irreducible (1PI) propagators and vertex functions is
generated by the renormalized effective action in the limit $k \to
0$. Thus the $k$-dependent action provides a smooth interpolation
between the bare (classical) action at an arbitrary ultraviolet (UV)
scale $\Lambda$ where no quantum fluctuations are taken into account
and the renormalized action in the IR.  For further details we refer
the interested reader to refs.~\cite{bergintro}, \cite{papp},
\cite{scha}, \cite{wett}.

The effective action is characterized by infinitely many couplings
multiplying all possible invariants which are consistent with the
symmetries of theory in question. In order to proceed one has to make
approximations. One possibility is the expansion of the effective
action in powers of the fields about an arbitrary field
$\Phi_0(x)$. The Taylor coefficients of this expansion
$\Gamma^{(n)}(x_1,\ldots ,x_n)$ then represent the full 1PI propagators and
vertices.

Alternatively one can expand the effective action in powers of
derivatives corresponding to an expansion in powers of the momenta
which we will pursue in this work.\footnote{ Which expansion pattern
for the effective action should be used depends on the underlying
theory. The derivative expansion will not be useful in situations
where the momentum-dependence of the correlation functions are
manifestly influenced by the interaction. } The invariants are in this
case classified by the number of the appearing derivatives and the
coefficients are functions of constant fields. For a one-component
scalar theory in $d$-dimensional coordinate space this expansion takes
the form
\be
\label{actionexp}
\Gamma [\Phi_c] =\int d^d x \left\{ -V(\Phi_c ) +\frac 12 Z \left( \Phi_c^2
\right)
\left( \partial_{\mu} \Phi_c \right)^2 + Y(\Phi_c^2) \left(
\left(\partial_{\mu} \Phi_c \right)^2
\right)^2 + ... \right\} \ .
\ee
The lowest-order term, the effective potential $V$, should be a convex
function, with no dependence on derivatives and can be rewritten as a
sum of all 1PI Green functions $\Gamma^{(n)}$ with $n$ external lines
carrying zero momenta. In other words the effective potential is the
generating functional of the zero-momentum 1PI Green functions. In the
following the expansion (\ref{actionexp}) will be truncated at second
order ${\cal O}(\partial^4)$ to derive a set of coupled partial
differential equations for the effective potential $V$ and
wavefunction renormalization $Z$ at the scale $k$.

\section{Application to the $O(N)$-model}
\label{section3}

In this section the RG method will be applied to the $O(N)$-symmetric
linear sigma model. We concentrate on the meson fields ($N-1$ massless
pions and one massive radial sigma mode) and work with an arbitrary
$N$ instead of $N=4$ fixed.

We employ the Euclidean metric due to the heat-kernel regularization
for the real part of the functional determinant\footnote{For notations
and details see e.g.~\cite{heat}}. This metric also allows for a
straightforward extension to finite temperature within the imaginary
time formalism as discussed in Section~\ref{finitetemperature}
\footnote{An implementation of the Wilsonian RG in the real-time
formulation can by found e.g.~in \cite{real} }.

At the scale $\Lambda$ in the ultraviolet region of the theory we
define the effective Lagrangian for the $O(N)$-model by
\be
{\cal L}_\Lambda = \f{1}{2} \left( \del_\mu \vec{\Phi} \right)^2 + V
(\vec{\Phi}^2)\ ;
\qquad V (\vec{\Phi}^2) =
\frac{\lambda}{4} \left( \vec{\Phi}^2 - \Phi_0^2 \right)^2
\ee
with the $N$-component vector $\vec{\Phi} = (\Phi_1, \Phi_2, \ldots ,
\Phi_N )$. The negative sign in the potential $V$ signals the broken
phase and a positive sign indicates the symmetric phase.

The field $\Phi_0$ denotes the minimum of the potential.  If this
field is constant only the lowest-order term (the effective potential)
of the derivative expansion of the effective action
Eq.~(\ref{actionexp}) can be computed. In the literature this
approximation is called the 'local potential approximation'.

Higher-order terms of the derivative expansion can be calculated with a
varying $x$-dependent background field $\Phi_0 (x)$ (see
e.g. \cite{morri}, \cite{cwet}, \cite{bona}). The expansion of the
effective action of the $O(N)$-symmetric model up to order ${\cal O}
(\partial^4)$ reads
\be
\label{actionon}
\Gamma [\vec{\Phi}] =\int d^d x \left\{ -V(\vec{\Phi}^2 ) +\frac 12
Z_1 \left( \vec{\Phi}^2 \right)
\left( \partial_{\mu} \vec{\Phi} \right)^2 + \f 12 Z_2 \left(
\vec{\Phi}^2  \right)
\left(\vec{\Phi} \partial_{\mu} \vec{\Phi} \right)^2
\right\} \ .
\ee
To this order there are two independent wavefunction renormalization
terms $Z_1$ and $Z_2$ for $N>1$ \footnote{The next order in the
derivative expansion would yield ten independent terms.}.

In the following we will sketch the derivation of the flow equation
including the wavefunction renormalization corrections. The
perturbative one-loop contribution to the effective action yields
formally a non-local logarithm which must be regularized. Proper-time
regularization is the most commonly used regularization for the real
part of the determinant and results in a finite local action (see
e.g.~\cite{ripk}). It was introduced by Schwinger and maintains a
finite sharp dimensionful proper-time cutoff~\cite{schw}. We replace
this sharp ultraviolet cutoff for the proper-time variable $\tau$ by a
smooth, $a$ $priori$ unknown, kernel or 'blocking function' $f_k$ in
the integrand which will be specified later.

After introducing a complete set of plane wave states in the heat kernel
\cite{nepo} this results in the following $d$-dimensional effective
action
\be
\label{gammaheat}
\Gamma [\vec{\Phi}] = -\frac 12 \int d^d x \int_0^{\infty} \frac{d \tau}{\tau} f_k
\int 
\frac{d^d p}{(2\pi)^d} \mbox{tr} \ e^{- \tau 
\left( p^2 - 2ip_{\mu} \partial_{\mu} - \partial^2
+V_{ij}'' (\Phi)\right)}
\ee
with the shorthand notation $V_{ij}'' (\Phi ) = \frac{\delta^2
V}{\delta\Phi_i \Phi_j}$ for the $(N \times N)$-matrix valued second
derivative of the $O(N)$-symmetric potential $V$ given by
\be
V_{ij}''= \lambda \left( \vec{\Phi}^2 -\Phi_0^2 \right) \delta_{ij} + 2
\lambda \Phi_i \Phi_j \ .
\ee
The trace in Eq.~(\ref{gammaheat}) runs over the $N$ fields and can be
computed analytically with standard techniques~\cite{fras},
\cite{jzuk}.  In the following we omit the index structure of the
potential term $V_{ij}''$ and follow the techniques outlined in
ref.~\cite{oles} which we generalize to an $O(N)$-symmetric
matrix-valued potential $V''$.

The expansion of $\exp \left[ -\tau \left( -2ip_\mu \partial_\mu
-\partial^2 + V'' \right) \right]$ in Eq.~(\ref{gammaheat}) in powers
of de\-ri\-va\-tives up to second order yields
\begin{eqnarray}
\label{lagordertwo}
\Gamma_2 &=& -\frac 12 \int\!\! d^d x \!\int\! \frac{d \tau}{\tau} f_k
\int 
\frac{d^d p}{(2\pi)^d} e^{- \tau p^2} \nonumber \\
&& \cdot \tr \left\{ \sum_{n=0}^{\infty} (-1)^n \frac{\left( V'' \right)^n
\tau^n}{n!}
+ \sum_{n=2}^{\infty} \frac{(-1)^{n-1} \tau^n}{n!} \sum_{k=0}^{n-1}
\left( V'' \right)^k
\partial^2 \left( V'' \right)^{n-1-k} \right. \nonumber \\
&& \left. \ \ \ -4 p_{\mu} p_{\nu} \sum_{n=3}^{\infty} \frac{(-1)^{n-1} 
\tau^n}{n!} \sum_{k=0}^{n-2} \left( V''\right)^k \partial_{\mu}
\sum_{l=0}^{n-2-k}
\left( V'' \right)^l \partial_{\nu} \left( V'' \right)^{n-2-l-k}
\right\}\ \ \ \ 
\end{eqnarray}
where all linear terms in the momenta $p_{\mu}$ are ignored since they
vanish anyhow after odd $p$-integration. The above expression can be
further simplified by a tedious but straightforward
calculation~\footnote{For technical details see the appendix
\ref{apptech}. }. All infinite sums converge and can be evaluated
analytically with careful attention of the non-commuting operator
order.

To improve readability we will use in the following the
abbreviations
\be
\label{massabb}
m^2_\sigma \equiv \lambda (3 \vec{\Phi}^2 - \Phi_0^2
)\quad\mbox{and}\quad m^2_\pi \equiv \lambda (\vec{\Phi}^2 - \Phi_0^2
)
\ee
and exploit the $O(d)$ symmetry of the momentum integration.

To order ${\cal O}(\partial^4)$ the
effective action splits into two parts:
\be
\Gamma_2 = \Gamma^{(0)} + \Gamma^{(2)}\nonumber
\ee
with $\Gamma^{(0)}$ denoting the one-loop potential contribution (no
derivatives)
\be
\label{gamma0}
\Gamma^{(0)} = - \frac 12 \int d^d x \int \frac{d \tau}{\tau}
f^{(0)}_k \int \frac{d^d p}{(2\pi)^d} e^{-\tau p^2} \left\{ 
e^{- \tau m_{\sigma}^2 } + (N-1) e^{-\tau m_{\pi}^2 } \right\} 
\ee
and $\Gamma^{(2)}$ the second-order contribution (containing two derivatives) 
\begin{eqnarray}
\label{gamma2}
\Gamma^{(2)} &=& \frac 12 \int d^d x \int \frac{d \tau}{\tau}
f^{(1)}_k \int \frac{d^d p}{(2\pi)^d} e^{-\tau p^2} \left\{ 
\left( \frac{\tau^2}{2} \partial^2 m_{\sigma}^2 -\frac{\tau^3}{3} 
\partial_{\mu} m_{\sigma}^2 \partial_{\mu} m_{\sigma}^2 \right) 
e^{-\tau m_{\sigma}^2 }  \right.\nonumber\\
&& \hspace*{4cm} +(N-1) \left( \frac{\tau^2}{2} \partial^2 m_{\pi}^2 -\frac{\tau^3}{3} 
\partial_{\mu} m_{\pi}^2 \partial_{\mu} m_{\pi}^2 \right) 
e^{-\tau m_{\pi}^2 } \nonumber \\
&& \qquad\quad-2 \left( \frac{2}{m_{\sigma}^2-m_{\pi}^2} \left( e^{-\tau
m_{\sigma}^2 } 
- e^{-\tau m_{\pi}^2 } \right)
+\tau \left( e^{-\tau m_{\sigma}^2 } + e^{-\tau m_{\pi}^2 } \right)
\right) \nonumber \\
&&  \hspace*{7.1cm}\left(
\frac{ ( \Phi_a \partial_{\mu} \Phi_a )^2}{ (\vec{\Phi}^2 )^2} -
\frac{ (\partial_{\mu} \Phi_a)^2}{ \vec{\Phi}^2} \right) \nonumber \\
&& - \frac{p^2 \delta^{\mu \nu}}{d} \left[ \left( \frac{2\tau^3}{3} 
\partial^2 m_{\sigma}^2 -\frac{\tau^4}{2} \partial_{\mu} m_{\sigma}^2 
\partial_{\mu} m_{\sigma}^2 \right) e^{-\tau m_{\sigma}^2} \right. \\
&&  \qquad\qquad + (N-1) \left( \frac{2 \tau^3}{3} \partial^2 m_{\pi}^2 
-\frac{\tau^4}{2} \partial_{\mu} m_{\pi}^2 \partial_{\mu} m_{\pi}^2
\right) 
e^{-\tau m_{\pi}^2 } \nonumber \\
&& \qquad\qquad + 2 \tau \left(  \frac{2}{m_{\sigma}^2-m_{\pi}^2} 
\left( e^{-\tau m_{\sigma}^2 } 
- e^{-\tau m_{\pi}^2 } \right)+  \tau \left( e^{-\tau m_{\sigma}^2 } 
+ e^{-\tau m_{\pi}^2 } \right) \right) \nonumber \\
&&  \hspace*{6.5cm}\left. \left.  \left( 
\frac{ (\partial_{\mu} {\Phi}_a)^2}{ \vec{\Phi}^2}
- \frac{ ({\Phi}_a \partial_{\mu} {\Phi}_a )^2}{ (\vec{\Phi}^2)^2} 
 \right) \right]
\right\} \nonumber \ .
\end{eqnarray}
For $N = 1$ this outcome for $\Gamma^{(2)}$ agrees with that
obtained in ref.~\cite{oles} using a similar procedure.

In order to extract the two different wavefunction renormalization
$Z_1$ and $Z_2$ of the effective action $\Gamma$ in
Eq.~(\ref{actionon}) one has to collect the terms in front of
$(\partial_\mu \vec{\Phi})^2$ and of $(\vec{\Phi} \partial_\mu
\vec{\Phi})^2$. To simplify the amount of work drastically we will use
the truncation to a uniform wavefunction renormalization which means
we neglect the field and momentum dependence of the wavefunction
renormalization. This approximation corresponds to considering only
the wavefunction renormalization constants at the minimum of the
potential $\Phi_0$ and neglecting their derivatives.

For $N>1$ one has to work with two different $Z_1$ and $Z_2$
corresponding to two anomalous dimensions.  These are related to the
wavefunction renormalization for the Goldstone modes $Z_{\pi}$ and the
radial mode $Z_{\sigma}$ via
\[
Z_{\pi} = Z_1 \quad \quad ; \quad \quad Z_{\sigma} = Z_1 +\vec{\Phi}^2 Z_2 \ .
\]
As will be seen later the
difference between the corresponding anomalous dimensions at the
scaling solution vanishes. We can further simplify the calculation by
neglecting $Z_2$. This approximation describes all the qualitative
features at the phase transition without loss of quantitative
predictive power.

Comparing the expressions in (9) and (10) with Eq.~(\ref{actionon}) 
we can now extract the effective potential contribution 
\be
\label{potential}
V = - \f {1}{2 (4\pi)^{d/2}} \int \frac {d \tau} {\tau}
\frac{1}{(Z \tau)^{d/2}} f^{(0)}_k \left\{
e^{-\tau m_\sigma^2} + (N-1) e^{-\tau m_\pi^2} \right\} 
\ee
and the $Z_1 \equiv Z$ wavefunction renormalization contribution
\ba
\label{wavefunction}
Z &=& -\f{1}{(4\pi)^{d/2}}\f{Z}{\Phi_0^2}\int \frac {d \tau} {\tau}
\frac{1}{(Z \tau)^{d/2}} \left[ \left( f^{(1)}_k - 2 f^{(0)}_k\right)\right. \\
&&\qquad\qquad
\left.\left\{ \tau \left( e^{-\tau m_\sigma^2} + e^{-\tau m_\pi^2} \right) +
\f {2}{m^2_\sigma - m^2_\pi}
\left( e^{-\tau m_\sigma^2} - e^{-\tau m_\pi^2} \right)\right\}\right]\nonumber\ ,
\ee
where momenta have been rescaled by the corresponding wavefunction
renormalization.

Note, that we have introduced two different blocking functions,
$f^{(0)}_k$ and $f^{(1)}_k$ to be discussed in the next Section. To
proceed we differentiate Eqs. (\ref{potential}), (\ref{wavefunction}) 
with respect to the scale $k$ which is introduced via the blocking
functions thus obtaining the desired flow equations for the potential
and wavefunction renormalization. In doing so one has to know the
derivative of the blocking functions $f^{(i)}_k$ $i=0,1,\ldots$ with
respect to $k$.

\subsection{The regulating blocking functions}

In this Section we sketch the way how to implement the regulating
blocking functions $f^{(i)}_k(\tau)$ in the proper-time
integrations\footnote{Other authors \cite{lia2} call the kernel $f_k
(\tau )$ also smearing function.}. They serve to cut off the diverging
momentum contributions to the flow equations~(see e.g.~\cite{oles,
lia2}). Going beyond the local potential approximation by including
higher contributions to the effective potential through the
wavefunction renormalization we have to deal with different powers of
momenta. In deriving flow equations for an $O(N)$-symmetric potential
we there encounter additional momentum integrals which force us to
modify the blocking functions. In fact, at each level of the gradient-
or derivative expansion new blocking functions must be taken into
account (cf.~e.g.~\cite{lia2}). In order to find a condition on the
structure of the blocking functions we compare the operator
regularization method to a momentum regularization method. In
ref.~\cite{papp} a differential equation for the blocking function in
the local potential approximation was found by comparing the operator
cutoff to a sharp momentum cutoff. This yields RG flow equations with
a characteristic logarithmic structure similar to the Wegner-Houghton
equation~\cite{wegn}. When the momenta are rescaled as $p^2 \to Zp^2$
we obtain a similar $d$-dimensional equation\footnote{A prime on
$f^{(i)}_k$ implies differentiation with respect to the scale $k$.}
\ba
\label{difffknull}
k f_k^{(0)\ \prime} & = & -\f {2}{\Gamma ( d/2 )} \left( \tau Z k^2 \right)^{d/2}
e^{- \tau Z k^2}\ .
\ea
The function $f_k^{(0)}$ regulates momentum integrals of the form
\be
I_0 = \int \f{d^d p}{(2\pi)^d} e^{-\tau Z p^2}\ .
\ee
and plays a similar role as the covariant Pauli-Villars cutoff.
Its derivation can be found in detail in
refs.~\cite{papp},\cite{scha}.  

The above expression must be modified when higher derivatives are taken
into account because one encounters higher powers in the momentum
integrals of the form e.g.
\ba
\label{i2}
I_2 = \int \f{d^d p}{(2\pi)^d} e^{-\tau Z p^2} p_\mu p_\nu & = & \f{\delta_{\mu\nu}}{d}
\int \f{d^d p}{(2\pi)^d} e^{-\tau Z p^2} p^2\ .
\ea
Comparing the sharp momentum cutoff version of $I_2$ with the
proper-time expression with a new blocking function $f^{(1)}_k$ and
by taking the derivative with respect to $k$ one obtains the 
following equation
\ba
k f^{(1)\ \prime} & = & - \f{4}{d \Gamma( d/2)} \left( \tau Z k^2 \right)^{d/2 + 1}
e^{- \tau Z k^2}\ .
\ea
Thus each power of $p^2$ in the numerator of the integrand
(cf.~Eq.(\ref{i2})) generates an additional power in the proper-time
variable $\tau$ thereby accelerating the convergence of the flow
equation. This result is identical to that in ref.~\cite{lia2}.

For completeness we quote the result for the blocking function
$f_k^{(2)}$ which regularizes momentum integrals of the form
\ba
\label{i4}
I_4 = \int \f{d^d p}{(2\pi)^d} e^{-\tau Z p^2} p_\mu p_\nu p_\lambda
p_\rho\ .
\ea
Although they are not needed in our approximation we will employ such
higher blocking functions when we analyze the influence of the blocking
functions on the convergence of the flow equations in
Sect.~\ref{infblocking}.

Comparing $f_k^{(2)}$ in the same manner with a sharp cutoff version yields the 
differential equation
\ba
k f_k^{(2)\ \prime} & = & - \f{8}{d (d+2) \Gamma(d/2)} \left( \tau Z k^2 \right)^{d/2 + 2}
e^{- \tau Z k^2}\ .
\ea

\subsection{The flow equations}
\label{section32}

With the general differential equation for the blocking function
$f^{(i)}_k$ at hand we can immediately obtain the flow equations. It
turns out that the lowest blocking function $f^{(0)}_k$ is sufficient
to regulate the potential equation alone\footnote{This blocking
function corresponds to the cutoff function of type (II) in
ref.~\cite{papp}.} (no derivatives) but not for higher momentum
contributions. The convergence of the solution of the flow equation is
very poor with $f^{(0)}_k$ when wavefunction contributions are taken
into account. This result agrees with that found in
\cite{papp},\cite{scha}. Thus in order to accelerate the convergence
from the very beginning we chose a blocking function $f^{(i)}_k$ with
$i>0$. We start with $i=1$ and replace in
Eqs.~(\ref{potential}),(\ref{wavefunction}) $f^{(0)}_k$ by
$f^{(1)}_k$. Performing now the convergent proper-time integration
this yields the flow equations
\be
\label{flowpotdim}
\partial_t V = 
S_d \f{k^d}{d}\left[ \f{1}{1+ 2v' + 4
\Phi^2 v''} + \f{N-1}{1+2v'} \right]
\ee
for the potential and  
\be
\label{flowzdim}
- \frac{1}{Z} \partial_t Z =  \left. \f{2 S_d}{\Phi^2 Zk^2}\f{k^d}{d} 
\left[ 1 + 
\frac{1}{ (1+4 \Phi^2 v'')^2 } + \frac{1}{2 \Phi^2 v''}
\left( 
\frac{1}{1+4 \Phi^2 v''} - 1 \right) \right] \right|_{{\Phi^2=\Phi_0^2}} 
\ee
for the wavefunction renormalization where we have introduced the
notations:
\be
\partial_t V \equiv k \f{\partial V}{\partial k}
\qquad,\qquad
v^{(i)} \equiv \f{V^{(i)}}{Zk^2}
\qquad\mbox{and}\qquad
S_d = \f{2}{\Gamma(d/2) (4\pi)^{d/2}}\nonumber\ .
\ee
A prime on the potential $V$ denotes differentiation with respect to
$\Phi^2$. Due to our uniform wavefunction renormalization
approximation (no derivatives of $Z$) we have to evaluate the flow
equation at the potential minimum $\Phi^2_0$.  The functions in
squared brackets describe the threshold behavior of massive modes and
are sometimes called 'threshold functions' \cite{wett}. Each massive
mode decouples from the evolution towards the infrared limit $k\to 0$
(see details in \cite{scha}). One advantage of our choice of the
blocking functions is that the threshold functions can be obtained
analytically. In Eq.~(\ref{flowpotdim}) one recognizes the persistent
contribution of the ($N-1$) massless Goldstone particles to the
evolution since $v'$ vanishes for the minimum of the potential (cf
also Eq.~(\ref{flowzdim})).

In the above flow equations we, in addition, perform a renormalization
group improvement thus going beyond the independent-mode
approximation. The improvement consists of replacing the bare
potential $V$ (and its derivatives) on the $rhs$ of the flow equations
with the running $k$-dependent counterpart $V_k$. Then higher graphs
such as daisy and super-daisy diagrams are taken into account. Thus we
substitute
\be
\label{masspotential}
m_\sigma^2 \equiv 2 V'_k + 4 \Phi^2 V''_k \qquad\mbox{and}\qquad 
m_\pi^2 \equiv 2 V'_k\ ,
\ee
which should not be confused with the abbreviation in
Eq.~(\ref{massabb}).  Of course, the potential $V_k$ is still a function
of the field $\Phi^2$ itself.  This type of nonlinear flow equations
has to be solved numerically on a grid as discussed in the next
Section.

\subsection{Finite temperature}
\label{finitetemperature}

Here we wish to discuss the generalization to finite
temperature. Within the Matsubara formalism this is easily
accomplished by replacing the $d$-dimen{\-}sional momentum integration
in Eqs.~(\ref{gamma0}) and (\ref{gamma2}) by a $(d-1)$--dimensional
integration and by introducing a discrete Matsubara summation with
bosonic frequencies $\omega_n = 2n\pi T$ in the zero-momentum
component.  An additional length scale (the inverse temperature) is
now introduced and besides quantum fluctuations additional thermal
fluctuations appear. At finite temperature the scale $k$ serves a
generalized IR cutoff for a combination of the three-dimen{\-}sional
momenta and discrete frequencies. The influence of the different
choices of blocking functions on the two types of modes can be found
in more detail in \cite{liao98}. We use in this work cutoff blocking
functions with a $d$-dimensional momentum variable.

In order to be consistent with the zero-temperature flow equations we
make the same approximations and use the same blocking function
$f_k^{(1)}$ at finite temperature. The finite-temperature extension of
Eqs.~(\ref{flowpotdim}) and (\ref{flowzdim}) with a uniform
wavefunction renormalization contribution is
\ba
\label{flowpotdimT}
\partial_t V &=& \f{Tk^3}{8(4\pi)} 
\left\{ \MB{0}{\f{m^2_\sigma}{Zk^2}}{\f{3}{2}} +
(N-1) \MB{0}{\f{m^2_\pi}{Zk^2}}{\f{3}{2}} \right\}\qquad 
\ea
where the following abbreviations for the
fi\-ni\-te\--tem\-per\-a\-ture threshold functions
\ba
\label{thresholdT}
\MB{p}{m^2}{\alpha} &=& \sum_{n=-\infty}^\infty
\f{(\omega^2_n)^p}{(1+\omega^2_n/k^2 + m^2)^\alpha}\ 
\ea
have been introduced together with 
the short-hand notation of Eq.~(\ref{masspotential}).
For the uniform finite-temperature wavefunction flow
equation we obtain accordingly
\ba
\label{flowzdimT}
\partial_t Z &=&-\f{Z}{\Phi_0^2}\f{Tk^3}{(4\pi)^{3/2}}
\left\{ \f{5}{2}\f{\Gamma(3/2)}{m^2_\sigma-m^2_\pi}
\left[ \MB{0}{\f{m^2_\sigma}{Zk^2}}{\f{3}{2}} - \MB{0}{\f{m^2_\pi}{Zk^2}}{\f{3}{2}}
\right]\right.\nonumber\\
&&\qquad\qquad\qquad\left.+ \f{5}{4}\f{\Gamma(5/2)}{Zk^2}
\left[ \MB{0}{\f{m^2_\sigma}{Zk^2}}{\f{5}{2}} + \MB{0}{\f{m^2_\pi}{Zk^2}}{\f{5}{2}}
\right]\right.\nonumber\\
&&\qquad\qquad\quad\left.+ \f{\Gamma(5/2)}{m^2_\sigma-m^2_\pi}\f{1}{k^2}
\left[ \MB{1}{\f{m^2_\pi}{Zk^2}}{\f{5}{2}} - \MB{1}{\f{m^2_\sigma}{Zk^2}}{\f{5}{2}}
\right]\right.\nonumber\\
&&\qquad\qquad\left.- \f{\Gamma(7/2)}{2 Zk^2}\f{1}{k^2}
\left[ \MB{1}{\f{m^2_\sigma}{Zk^2}}{\f{7}{2}} + \MB{1}{\f{m^2_\pi}{Zk^2}}{\f{7}{2}}
\right]\right\}.
\ea

It is possible to calculate the zero-temperature limit of
the finite-tem\-pe\-ra\-ture flow equation (\ref{flowpotdimT}) and
(\ref{flowzdimT}) analytically by using the zero-temperature limit 
of the threshold functions\footnote{see e.g. ref.\cite{scha}.}:
\be
\MB{0}{m^2}{\alpha} \to \f{k}{\pi T}
\f{2\cdot4\cdots(2\alpha-3)}{1\cdot3\cdots(2\alpha -2)}
\f{1}{(1+m^2)^{(2\alpha-1)/2}}
\ee
and
\be 
\MB{1}{m^2}{\alpha} \to \f{k^3}{2\pi T}
\f{2\cdot4\cdots(2\alpha-5)}{1\cdot3\cdots(2\alpha -4)}\f{1}{(\alpha-1)}
\f{1}{(1+m^2)^{(2\alpha-3)/2}}\ .
\ee

In the next section we discuss the numerical solution of these
equations for the broken phase.

\section{Results}
\label{section4}

\subsection{Numerical implementation}

We discretize the field $\Phi^2$ for a general potential term
$V(\Phi^2 )$ on a grid and numerically solve the coupled set of flow equations
(\ref{flowpotdim}),(\ref{flowzdim}) for zero temperature and
(\ref{flowpotdimT})-(\ref{flowzdimT}) for finite temperature
at each grid point.

The flow equations incorporate derivatives of $V$ with respect to the
field $\Phi^2$ up to second order. Thus further differentiation of the
flow equation with respect to the fields increases the order of
derivatives of $V$. Following the method described in \cite{berg} we
calculate the first- and second derivative of the flow equation for
the potential with respect to $\Phi^2$ and thus generate derivatives
of $V$ up to fourth order. The unknown derivatives $V'''$ and $V''''$
are determined by matching conditions at intermediate grid points in
order to close the highly coupled system of differential
equations. This matching condition is based on a Taylor-expansion of
$V'$ and $V''$ up to forth order at each grid point $\Phi^2_i$ and it
requires continuity of these expansions at intermediate adjoining grid
points.

For $n$ grid points one obtains $2n-2$ equations for the $2n$ unknown
derivatives $V'''$ and $V''''$. The missing two conditions of the
boundary grid points are determined by an expansion of the third
derivatives. This results in an $2n$-dimensional algebraic coupled set
of flow equations for $V'$ and $V''$ which can be solved by a
fifth-order Runge Kutta algorithm with adaptive step size
\cite{press}.

With the grid algorithm described above the flow equations are solved
starting the evolution in the broken phase deep in the UV region at a
given $k = \Lambda$ where we use a tree-level parameterization of the
potential with only two initial parameters. For the results presented
below the values $\Lambda = 800$ MeV, $\Phi^2_{0,\Lambda}= (140 $
MeV$)^2$, $\lambda_{\Lambda} = 49$ and $Z_\Lambda = 0.8$ have been
chosen.

\begin{figure}[!htbp]
\vspace{-0.1in}
\centerline{\hbox{
\psfig{file=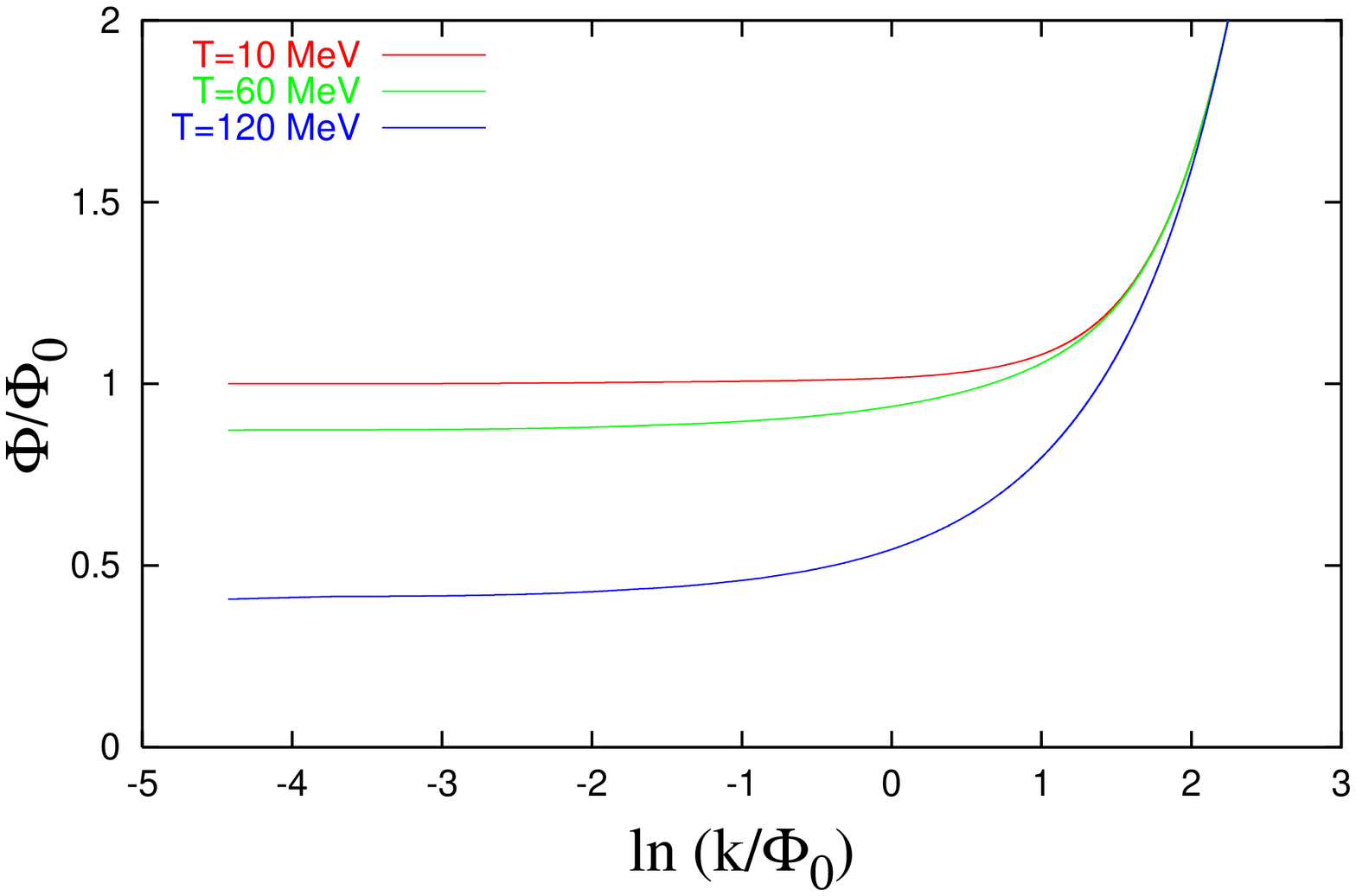,height=7cm,width=6cm,angle=-90}
\psfig{file=space.ps,width=0.05cm,angle=-90}
\psfig{file=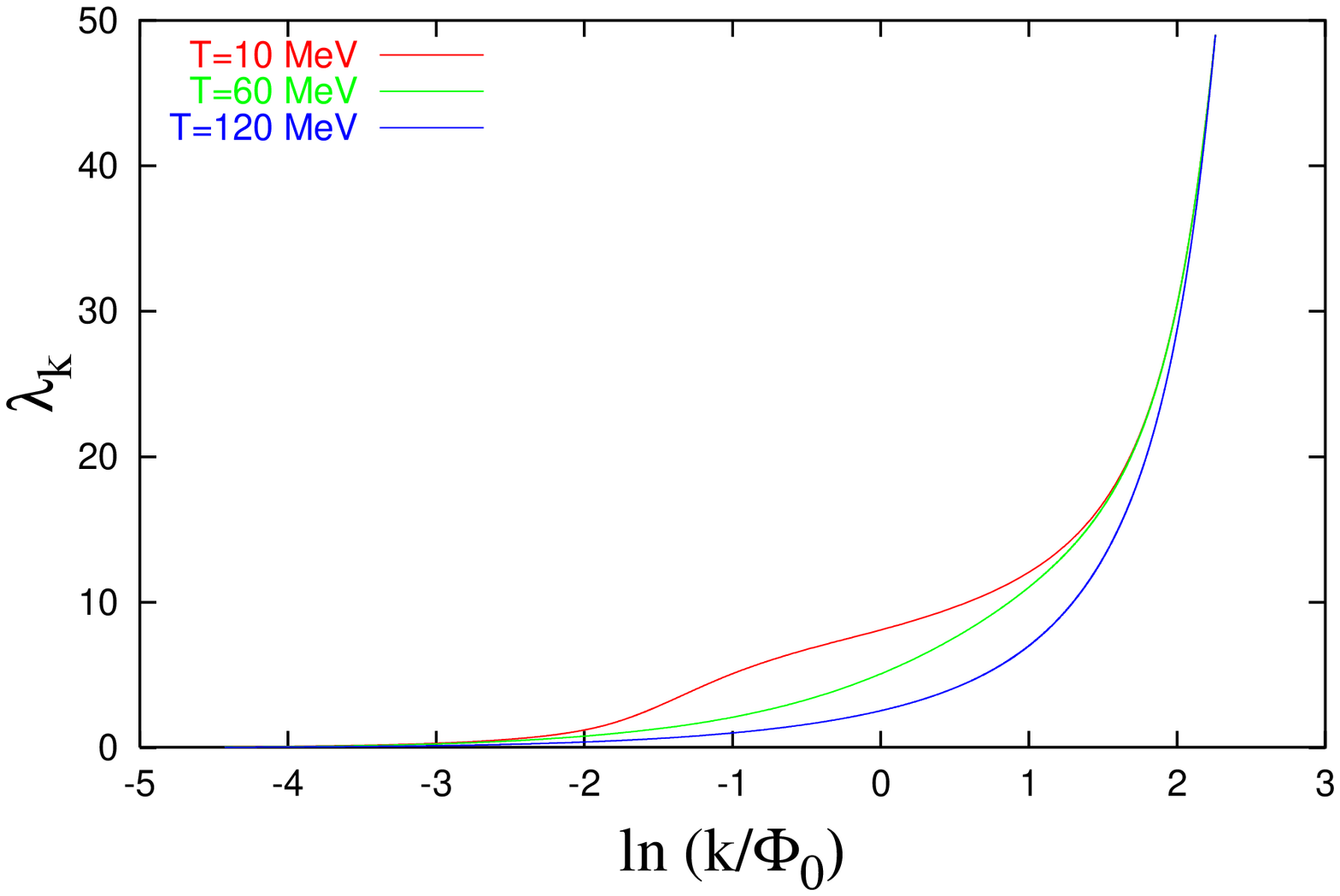,height=7cm,width=6cm,angle=-90}
}}
\centering
\parbox{12cm}
{\caption{\label{figkappaandlambdaT} The $k$-evolution of the
minimum of the full potential (left panel) and of the quartic coupling 
$\lambda_k$ (right panel) for various temperatures.}}
\end{figure}

The evolution with respect to the scale $k$ of the dimensionful
minimum of the potential $\Phi_{0,k}$ in units of $\Phi_{0} (T=0)$ and
the quartic coupling $\lambda_k$ for $N=4$ is shown in
Fig.~\ref{figkappaandlambdaT} for different temperatures without
wavefunction renormalization corrections\footnote{Including
wavefunction corrections does not change the shown curves
significantly.}. The initial values at the UV scale $\Lambda$ are
fixed at $T=0$ and are kept constant for all temperatures. This, in
principle, limits the predictive power of the high-temperature
behavior of the calculated quantities. We have checked that for the
range of temperatures relevant for the phase transition the results
are insensitive to temperature variations in $\Lambda$ within
reasonable limits.

For all temperatures below $T_c$ the minimum $\Phi_{0,k}/\Phi_{0}
(T=0)$ (left panel) decreases as a function of $k$ and settles down to
a constant for very small $k$-values where we can stop the evolution
to obtain the renormalized vacuum expectation value ($\Phi_0$)
(cf.~\cite{tetr}). Which is the physical order parameter for the
$O(N)$ symmetry phase transition.

The quartic coupling $\lambda_k$ which in four dimensions is a
dimensionless quantity shows a logarithmic running and tends to zero
for $k\to 0$ as can be seen in the right panel of
Fig.~\ref{figkappaandlambdaT}. For $N>1$ there are $(N-1)$ massless
pions in the spontaneously broken phase. Below the phase transition
they never decouple from the evolution and thus always contribute to
the logarithmic running resulting in an infrared-free theory. The
undulation in the curve for small temperature ($T=10$ MeV) is a
numerical artefact and stems from the finite grid spacing.

\subsection{The potential and the quark condensate}

\begin{figure}[!htb]
\centering
\vspace{-0.1in}
\psfig{file=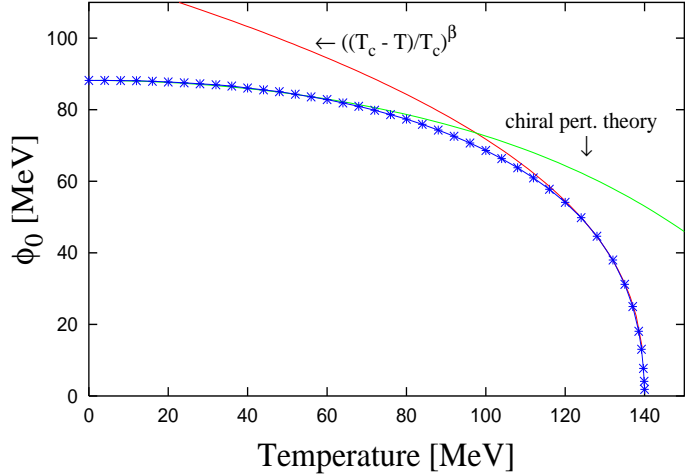,width=2.6in,height=3.5in,angle=-90}
\parbox{12cm}
{\caption{\label{figvev} The order parameter for the $O(4)$-model as
a function of temperature. A comparison with the chiral perturbation
theory expansion for low temperatures and with the Widom scaling
function in the vicinity of the critical temperature is also shown.}}
\end{figure}

In the infrared the order parameter $\Phi_0$ signals a
second-order phase transition (see Fig.~\ref{figvev}).
The vacuum expectation value is obtained as the dimensionful minimum
of the full effective potential with respect to the field, including
wavefunction renormalization contributions. For low temperatures we
compare our result for $N=4$ with a three-loop chiral perturbation
theory expansion up to order ${\cal O} \left( (T/f_\pi)^8 \right)$
\cite{chpt} and find perfect agreement up to $T \approx 45$ MeV. 
The critical temperature is a non-universal quantity and therefore
depends on the initial UV parameters.  Our value of $T_c \approx 140$
MeV is higher than the value $T_c\approx 100$ MeV of ref.~\cite{wett}
also obtained for a full potential analysis with the exact
renormalization group including wavefunction renormalization
contributions. This discrepancy in the value of $T_c$ might be due to
the different implementations of the threshold functions in the flow
equations.  In the vicinity of $T_c$ we see a scaling behavior of the
order parameter with a critical exponent $\beta \approx 0.4$ also
depicted in Fig.~\ref{figvev}.

\begin{figure}[!htbp]
\vspace{-0.1in}
\centerline{\hbox{
\psfig{file=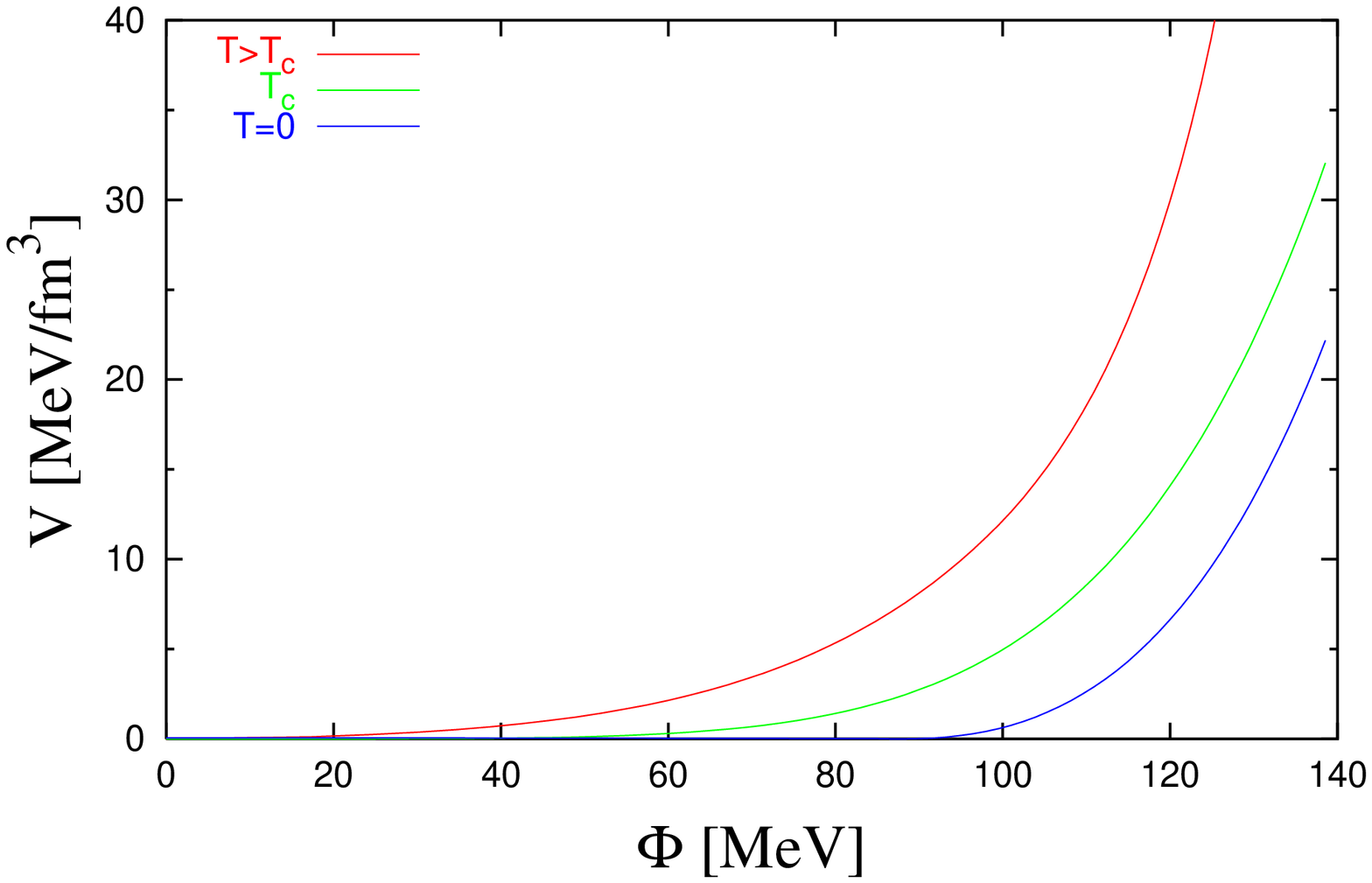,height=7cm,width=6cm,angle=-90}
\psfig{file=space.ps,width=0.02cm}
\psfig{file=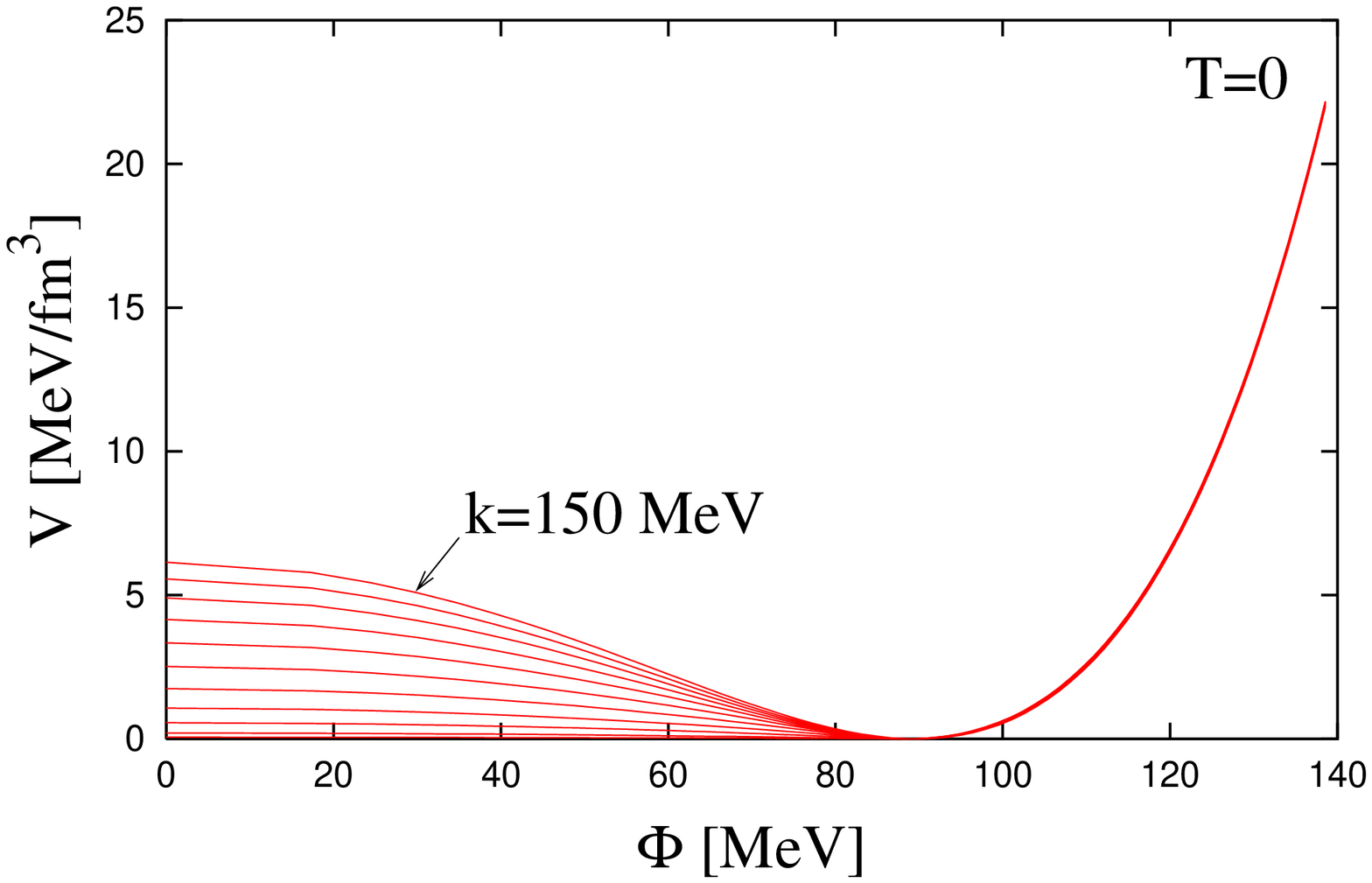,height=7cm,width=6cm,angle=-90}
}}
\centering
\parbox{12cm}
{\caption{\label{figpot} The dimensionful convex potential for
different temperatures (left panel) and the $k$-evolution of the dimensionful
potential at $T=0$ (right panel) towards zero.}}
\end{figure}

In our analysis we are not restricted to universal quantities such as
the minimum of the potential. Indeed we do know the entire potential
as function of the field $\Phi$ which is shown in Fig.~\ref{figpot}
(left panel) in the $k\to0$ limit for various temperatures ($T=0$,
$T=T_c$ and $T>T_c$). Here one again observes a second-order phase
transition. Numerically we have to stop the evolution at a very small
but finite $k$-value which slightly influences the final shape of the
potential. In fact, the potential at $k=0$ corresponds to the
free-energy density which should be a flat function around the origin
due to convexity. For a finite $k$-value we always obtain the
well-known Mexican hat form for the broken phase (right panel).  The
evolution towards the Maxwell construction (\cite{polo}) at $T=0$ can
explicitly be seen.

\section{Critical behavior}
\label{section5}

Within the RG approach we can directly link the zero-temperature
physics to the universal behavior in the vicinity of the critical
temperature. We will restrict ourselves to temperatures $T \le T_c$ in
this section but there is in general no restriction to the broken
phase.

In order to investigate the critical regime of the system it turns out
that the use of rescaled quantities is convenient since the scale $k$
is no longer present explicitly in the evolution equations. At the
critical temperature only properly rescaled quantities asymptotically
exhibit scaling and the evolution of the system is purely
three-di\-men\-sion\-al. This is the well-known dimensional reduction
phenomenon (see e.~g.~ref.~\cite{liaostrick}) and is exhibited in the
structure of the above mentioned threshold functions. It is also the
reason for the fractional power in the threshold functions at finite
temperature in Eqs.~(\ref{flowpotdimT})-(\ref{flowzdimT}). For the
dimensionally reduced system at high temperature ($T\gg k$) only the
bosonic zero-mode Matsubara frequencies are important while all the
other modes decouple completely from the evolution.

We define the rescaled (dimensionless) quantities as follows
\be
\label{defphi}
{\phi}^2 := k^{-(d-2+\eta)} \Phi^2
\ee
where we have introduced the anomalous dimension 
\be
\eta = -\frac{\partial}{\partial t} \ln Z_k .
\ee
As already mentioned in Sect.~\ref{section3} there are in general two
anomalous dimensions for the $O(N)$-model, one for the radial mode and
another one for the Goldstone bosons. Due to our uniform wavefunction
renormalization approximation we work with only one $\eta$\footnote{At
the scaling solution both anomalous dimensions become degenerate anyway.}.

In this section all results are obtained by taking the corrections due
to the anomalous dimension $\eta$ explicitly into account. In order to
deal with dimensionless flow equations we additionally rescale the
potential via
\be
u({\phi}^2) := k^{-d } V({\phi}^2)\ .
\ee
All dimensionless quantities are denoted in the following by small
letters.

From Eqs.~(\ref{flowpotdim})-(\ref{flowzdim}) we obtain the following
properly rescaled dimensionless flow equations in $d$ dimensions
\begin{eqnarray}
\label{flowpotcrit}
\partial_t u({\phi}^2) &=& - d u +(d-2+\eta ){\phi}^2
u' +
\frac{S_d}{d} \left[ \frac{1}{1+2u' +4{\phi}^2 u''}
+\frac{N-1}{1+2u'}
\right]\ \ 
\end{eqnarray}
\begin{eqnarray}
\label{flowzcrit}
- \frac{1}{Z} \partial_t Z &=&  \left. \frac{2 S_d}{{\phi}^2 d} 
\left[ 1 + 
\frac{1}{ (1+4 {\phi}^2 u'')^2 } + \frac{1}{2 {\phi}^2 u''}
\left( 
\frac{1}{1+4 {\phi}^2 u''} - 1 \right) \right] \right|_{\phi=\phi_0} \ ,
\end{eqnarray}
where the primes on the potential $u$ indicate differentiation with
respect to ${\phi^2}$. Since no scale $k$ appears explicitly in the
rescaled equations they are in a scale-independent form. Due to the
dimensional reduction phenomenon we do not need to use the finite
temperature flow equations at the critical temperature. It is
sufficient to employ the three-dimensional zero-temperature equations
in oder to investigate the critical regime of the phase transition.

\subsection{The scaling solution and fixed points}

We solve the set of coupled equations (\ref{flowpotcrit}),
(\ref{flowzcrit}) for arbitrary $N$ with the same initial values at
$t=0$\footnote{For the rescaled system no explicit value for the UV
cutoff $\Lambda$ is necessary.}. We arbitrarily set $\lambda = 0.5$
which is anyway an irrelevant quantity and finetune only
$\left. {\phi_0^2} \right|_{t=0}$ at $t=0$ in order to find a
$k$-independent solution, the so-called 'scaling solution' of the flow
equations. A second-order phase transition involves an infrared fixed
point of the renormalization group. Therefore the physics close to the
phase transition is scale invariant and the critical behavior should
be described by a scale-independent solution.

\begin{figure}[!htbp]
\centering
\vspace{-0.1in}
\psfig{file=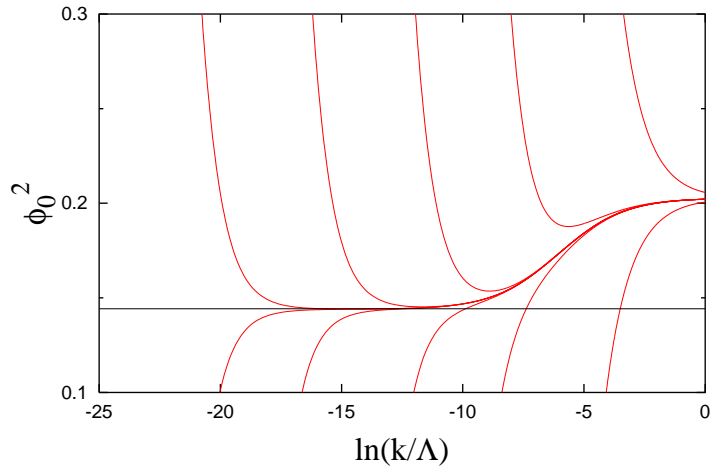,width=2.5in,angle=-90}
\parbox{12cm}
{\caption{\label{figscaling} The evolution of the dimensionless
minimum $\phi_0^2$ as function of the ``flow time'' $t=\ln(k/\Lambda)$
for $N=4$ and different initial values at $t=0$ around the critical
value $\left. {\phi_0^2} \right|_{t=0}^{cr}$ (see text for
details). }}
\end{figure}

For an initial value around the critical $\left. {\phi_0^2}
\right|_{t=0}^{cr}$ at the UV scale $t=0$ this is indeed the case and 
the evolution towards the infrared $t\to -\infty$ reaches almost the
scaling solution where the $t$-dependence vanishes and ${\phi_0^2}$
tends to a constant (fixed point) value ${\phi_0^2}^*$. This feature
is demonstrated in Fig.~\ref{figscaling} where the evolution of
${\phi_0^2}$ versus the ``flow time'' $t = \ln (k/\Lambda)$ is shown for different
initial values at the UV scale $t=0$.

For a starting value near the critical value $\left. {\phi_0^2}
\right|_{t=0}^{cr}$ the evolution deviates either towards the
spontaneously broken (${\phi_0^2} \neq 0$) or the symmetric
(${\phi_0^2} = 0$) phase in the infrared as shown in
Fig.~\ref{figscaling}.  Due to the rescaling of the dimensionful
minimum $\Phi_0^2$ (cf Eq.~(\ref{defphi})) which in the infrared
tends to a constant value or zero during the evolution with respect to
the scale $k$, the dimensionless minimum ${\phi_0^2} = \Phi^2_0/k^2$
diverges for $k \to 0$ for the broken phase as can be seen in 
Fig.~\ref{figscaling}.

In the further course of the evolution, for a initial value very close
to the critical one at $t=0$, the scaling regime is approached
where the system spends a long ``time'' before it finally
deviates from it again. The scaling solution for the minimum of the
potential is denoted by a straight line in Fig.~\ref{figscaling}.

Not only ${\phi_0^2}$ but all other quantities reach the scaling
solution around $t \approx -10$ and leave the critical trajectory at $t
\approx -18$. The time that the system spends on this scaling solution can
be rendered arbitrarily long by appropriate finetuning of the initial
values at the UV cutoff scale i.e.~at $t=0$.  In order to produce such a
high precision as shown in the Figure one has
to know the initial starting value up to $13$ digits.

\begin{figure}[htbp!]
\centering
\vspace{-0.1in} 
\psfig{file=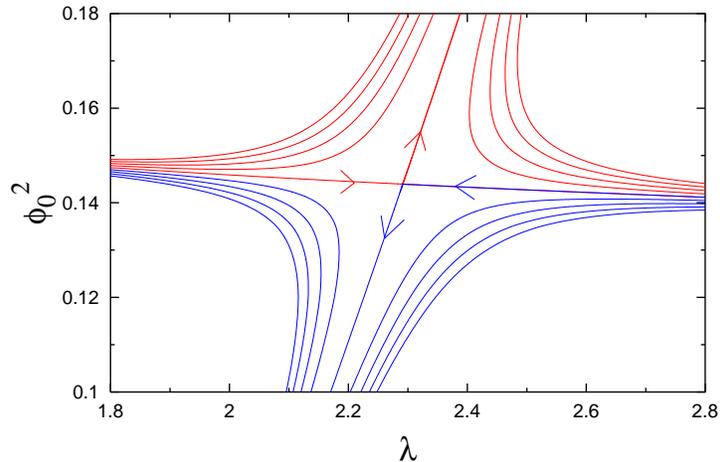,width=2.5in,angle=-90}
\parbox{12cm}
{\caption{\label{figfixpoint} The mixed fixed point for $N=4$.}}
\end{figure}

Besides of the trivial high-temperature and low-temperature fixed
points the $O(N)$-symmetric model exhibits a nontrivial mixed fixed
point inherent in the flow equations.  The region around this fixed
point is mapped out in Fig.~\ref{figfixpoint} where the rescaled
dimensionless quantities ${\phi_0^2} $ and $\lambda $ are plotted for
$N = 4$ and for different initial values during the evolution towards
the infrared. The arrows in the Figure characterize the flow
w.r.t.~the scale $t$ towards zero.

In general, there are two relevant physical parameters for the $O(N)$
model which must be adjusted to bring the system to the critical fixed
point according to universality class arguments. Consider for example
the ferromagnetic spin Ising model with a discrete symmetry $Z_2 =
O(1)$ where the relevant parameters are the temperature and external
magnetic field.  Since we work in this section in the chiral limit (no
external sources or masses) only one rele{\-}vant eigenvalue from the
linearized RG equations is left, the temperature\footnote{The
temperature-like relevant variable is also called thermal scaling
variable.}. This can be seen in Fig.~\ref{figfixpoint}. The quantity
${\phi_0^2}$ is the relevant variable because repeated renormalization
group iterations (which are the discrete analog to the continuous
evolution with respect to $t \to -\infty$) drive the variable away
from the fixed-point value while $\lambda$ is the irrelevant variable
and is iterated towards the fixed point if the initial values are
chosen sufficiently close to the fixed point. Thus one has a
one-dimensional curve of points attracted to the fixed point, the
so-called 'critical surface' where the correlation length diverges.

One nicely sees that the critical surface (line) separates both phases. When we
choose initial values near the critical line the systems spends a long
time near the critical point. For example starting with
${\left.\phi^2_0\right|_{t=0}} > \left.\phi^2_0\right|^{cr}_{t=0}$ and
$\lambda$ finite, the evolution tends ultimately to the
zero-temperature fixed point corresponding to very large values of
${\phi_0^2}$.  On the other hand, starting below the critical line the
evolution of ${\phi_0^2}$ tends towards the high-temperature fixed
point corresponding to small values of ${\phi_0^2}$. Only for exactly
${\left.\phi^2_0\right|_{t=0}} = \left.\phi^2_0\right|^{cr}_{t=0}$ the
renormalization group trajectory flows into the critical mixed fixed
point, which means that the large distance behavior at the critical
point is the same as that of the fixed point.

Here the predictive power of the nonperturbative approach becomes
visible: During the evolution near the scaling solution the system
loses memory of the initial starting value in the UV and the effective
three-dimensional dynamics near the transition is completely
determined by the fixed point and hence independent of the details of
the considered microscopic interaction at short distances.

The equation of state drives the potential away from the critical
temperature. As a result after the potential has evolved away from the
scaling solution its shape is independent of the choice of the initial
values such as e.g.~$\lambda$ for the classical theory at the UV,
signaling universal behavior near the cri{\-}tical point.

\subsection{Beta functions}

\begin{figure}[ht!]
\centering
\vspace{-0.1in}
\psfig{file=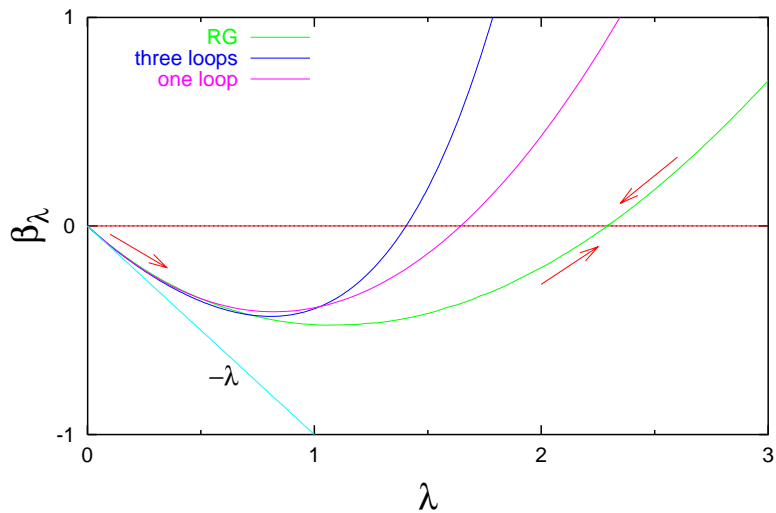 ,width=7cm , angle=-90}
\parbox{12cm}
{\caption{\label{figbetalambda}
The beta function $\beta_\lambda (\lambda)$ for the $O(4)$-model. The
RG result is compared to an one- and three-loop $\eps$-expansion for
three dimensions. The arrows denote the flow w.r.t. the scale $k$
towards the infrared.}}
\end{figure}

In Fig.~\ref{figbetalambda} the beta function $\beta_\lambda \equiv
\partial_t \lambda$ is presented as a function of $\lambda$. For $N=4$ it has 
two zeros at $\lambda^* =0$ and $\lambda^*= 2.28$ corresponding to
two fixed points. The position of the fixed points depends on the
particular form of the RG flow equation encoded in the choice
of the blocking function $f^{(i)}_k$. Physical results, however, such as
critical exponents must not depend on a particular choice.

\begin{figure}[htbp!]
\vspace{-0.1in}
\centerline{\hbox{
\psfig{file=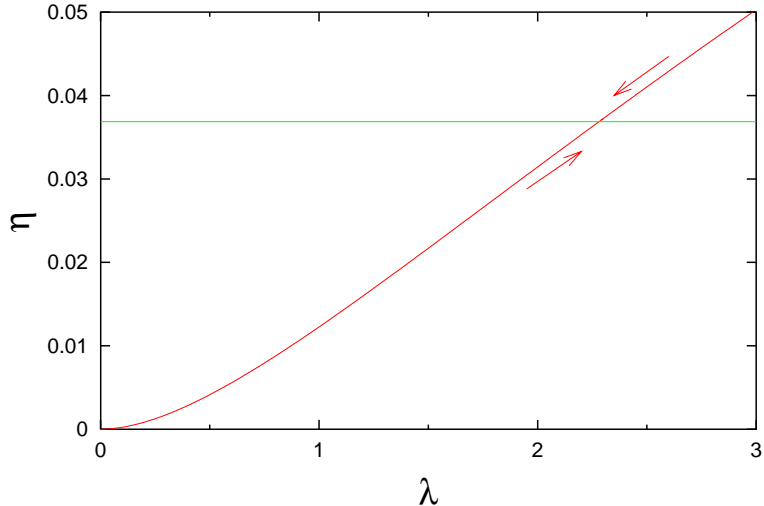,width=7cm,angle=-90}
}}
\centering
\parbox{12cm}
{\caption{\label{figbetakandeta} The anomalous dimension $\eta$ as a 
function of the quartic coupling $\lambda$. The flat line indicates $\eta$ 
at the physical vacuum.}}
\end{figure}

The anomalous dimension $\eta$ is nonzero at the nontrivial fixed
point as shown in Fig.~\ref{figbetakandeta}. In general it is
difficult to calculate the zeros of the beta function since this
requires knowledge of the physics beyond perturbation theory. The
ultraviolet-stable but infrared-unstable fixed point at $\lambda^*=0$
is called the Gaussian fixed point because the fixed-point Hamiltonian
yields a (simple) Gaussian partition function. The properties of this
fixed point depend on whether the dimension $d$ is greater or less
than 4. When $d$ is (slightly) smaller than 4 $\lambda$ is (slightly)
relevant at the Gaussian fixed point and thus can flow to another
nearby fixed point at $\lambda^* \approx 2.28$. This point is called
the Wilson-Fisher fixed point and the flow with respect to $k$ is
indicated in Fig.~\ref{figbetalambda} by arrows. Thus in the limit
$k\to 0$ the coupling $\lambda_k$ is driven to this infrared-stable
fixed point because the derivative of the beta function is positive.

\addtocounter{figure}{1}
\newcounter{figeins}
\setcounter{figeins}{\thefigure}

We can also verify the large-$N$ limit: The fixed point $\lambda^*$
tends to zero for $N\to\infty$ merging with the Gaussian fixed point (see
Table~{\ref{tab1}} and Fig.~{\thefigeins})~\cite{schn}. When the two
fixed points are sufficiently close to each other it is then possible
to deduce universal properties at one fixed point in terms of those at
the other one.

\begin{table}
\noindent
\vspace*{2ex}
\begin{minipage}{0.45\textwidth}{
{\hbox{
\psfig{file=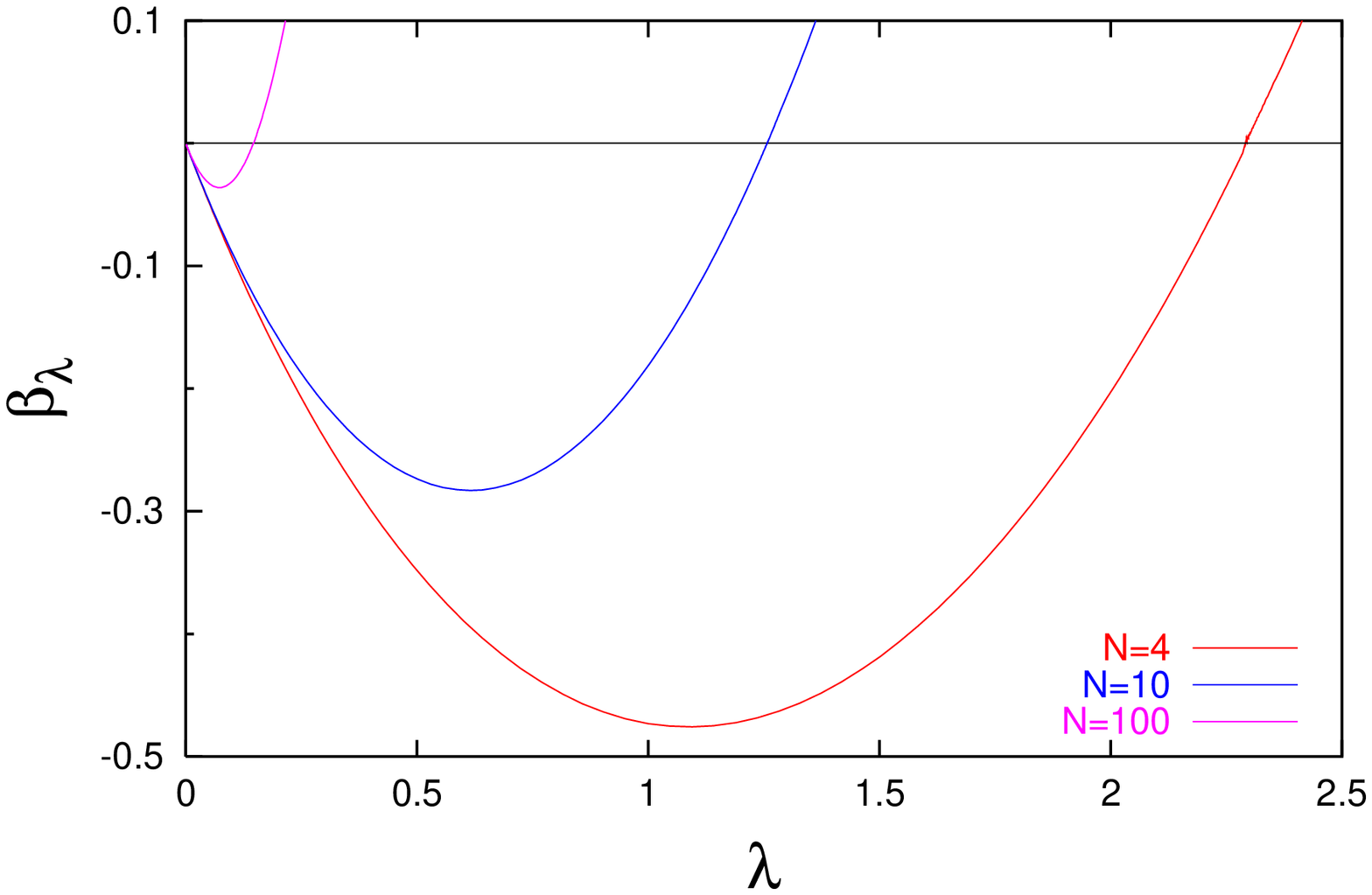,height=7cm,width=6cm, angle=-90}
}}}

\hspace{0.5cm}
\parbox{1.1\textwidth}{
Figure {\thefigure}: The beta function $\beta_\lambda (\lambda)$ for
various values of $N$ demonstrating the large-$N$ behavior. }
\end{minipage}\hfill
\begin{minipage}{0.5\textwidth}{
\begin{center}
\begin{tabular}{|l|c|c|c|c|c|}
\hline
 & ${\phi_0^2}^*$ & $\lambda^*$  \\
\hline
N=1   & 0.062     & 3.20      \\
N=2   & 0.087     & 2.88      \\ 
N=3   & 0.115     & 2.58      \\
N=4   & 0.144     & 2.29      \\
N=10  & 0.336     & 1.26      \\
N=100 & 3.369     & 0.15      \\
\hline
\end{tabular} 
\end{center}
\begin{center}
\parbox{0.85\textwidth}{
\caption{\label{tab1} In three-\-di\-men\-sions the Wil\-son\--Fish\-er
fix\-ed point tends towards zero in the large-$N$ limit.}}
\end{center}}
\end{minipage}
\end{table}

Universal quantities can be calculated by perturbation theory at fixed
dimension or by the $\eps$-expansion where the parameter $\eps = 4-d$
is related to the number of dimension of the system. The computation
of the critical exponents via the $\eps$-expansion beyond order ${\cal
O} (\eps)$ requires knowledge of complicated higher-order terms in the
perturbative renormalization group equations. It is believed that the
$\eps$-expansion (including the assumption of Borel summability) is a
good approximation scheme if a sufficient number of higher terms are
taken into account. At the end of the calculation one has to set
$\eps$ equal to one (or two) to obtain results of direct physical
significance. One should bear in mind, however, that for a given case
the $\eps$-expansion might be an asymptotic expansion giving eventually
inaccurate results if truncated at too high an order.

In Fig.~\ref{figbetalambda} we compare the beta function
$\beta_\lambda$ for $N=4$ obtained numerically within our
nonperturbative RG approach with $\eps$-expansions for different loop
orders.

The $\eps$-expansion for an $O(N)$-symmetry up to three-loop order
\cite{zinn} is given by (for $\eps = 1$)
\begin{eqnarray*}
\beta_{\lambda} &=& -g +\frac{(N+8)}{6}g^2
-\frac{(3N+14)}{12}g^3 \\
&&+\frac{33N^2+922N+2960+96(5N+22)\zeta (3)}{12^3} g^4 +O(g^5)
\end{eqnarray*}
with $\zeta(3) = 1.202057$ and the short-hand notation
\begin{eqnarray*}
g &=& \frac{12\lambda}{(4\pi)^{\frac 32} \Gamma \left( \frac 32
\right)}\ .
\end{eqnarray*}

At three-loop order we find agreement with our result up to
$\lambda \approx 0.55$ while the one-loop order deviates from our
result around $\lambda \approx 0.45$. Of course, the
$\eps$-expansion, despite of a failure from a numerical point of view,
provides an unambiguous classification of the fixed point consistent
with our nonperturbative RG results.

\subsection{Critical exponents and scaling relations}

In the vicinity of the critical point one obtains a scaling behavior
of the system which is governed by critical exponents. They
parameterize the singular behavior of the free energy near the phase
transition.  Altogether there are six critical exponents
$\alpha$, $\beta$, $\gamma$, $\delta$, $\nu$ and $\eta$ for the
$O(N)$-model but only two of them are independent due to four
scaling relations:
\begin{eqnarray}
\label{scalingrelation}
\alpha &=& 2 - d \nu \nonumber\\
\beta &=& \frac{\nu}{2} (d -2 +\eta ) \nonumber\\
\gamma &=& (2-\eta )\nu \\
\delta &=& \frac{d+2-\eta }{d-2+\eta } \ .\nonumber
\end{eqnarray}
In general, these universal critical exponents depend only on the
dimensionality of the system $d$ and its internal symmetry. The
exponents $\eta$ and $\delta$ describe the behavior of the system
exactly at the critical temperature and the remaining ones
parameterize the system in the region of the fixed point around $T_c$
($T\neq T_c$).

In this work we have calculated the critical exponents $\eta$,
$\beta$, $\nu$ and $\delta$ in three-dimensions numerically for
different values of $N$ and could verify the scaling
relations~(\ref{scalingrelation}) with deviations typically being less
than 0.1~$\%$. The smallness of the anomalous dimension $\eta$ (see
Table~2) justifies the use of the derivative expansion of the
effective action.

The critical exponent $\beta$ parameterizes the behavior of the
spontaneous magnetization or the order parameter for the broken phase
$(\left. {\phi_0^2} \right|_{t=0} > \left. {\phi_0^2} \right|_{t=0}^{cr} )$ 
in the vicinity of the critical temperature (see Fig.~\ref{figvev}). 
It is here numerically obtained as the slope of the
logarithm of ${\phi_0^2}$ as a function of $\ln(\left. {\phi_0^2}
\right|_{t=0} - \left. {\phi_0^2} \right|_{t=0}^{cr} )$ for very small
arguments. The difference $(\left. {\phi_0^2} \right|_{t=0} -
\left. {\phi_0^2} \right|_{t=0}^{cr} )$ is a measure of the distance
from the phase transition irrespective of any given value
$\lambda_{t=0}$ at the UV scale.  If $\left. {\phi_0^2} \right|_{t=0}$
is interpreted as a function of temperature this difference is
proportional to the deviation from the critical temperature
i.e. proportional to $(T - T_c)$. Thus the quantity $\left. {\phi_0^2}
\right|_{t=0}^{cr}$ defines the critical temperature in
three-dimensions.

\begin{figure}[!htb!]
\centering
\vspace{-0.1in}
\psfig{file=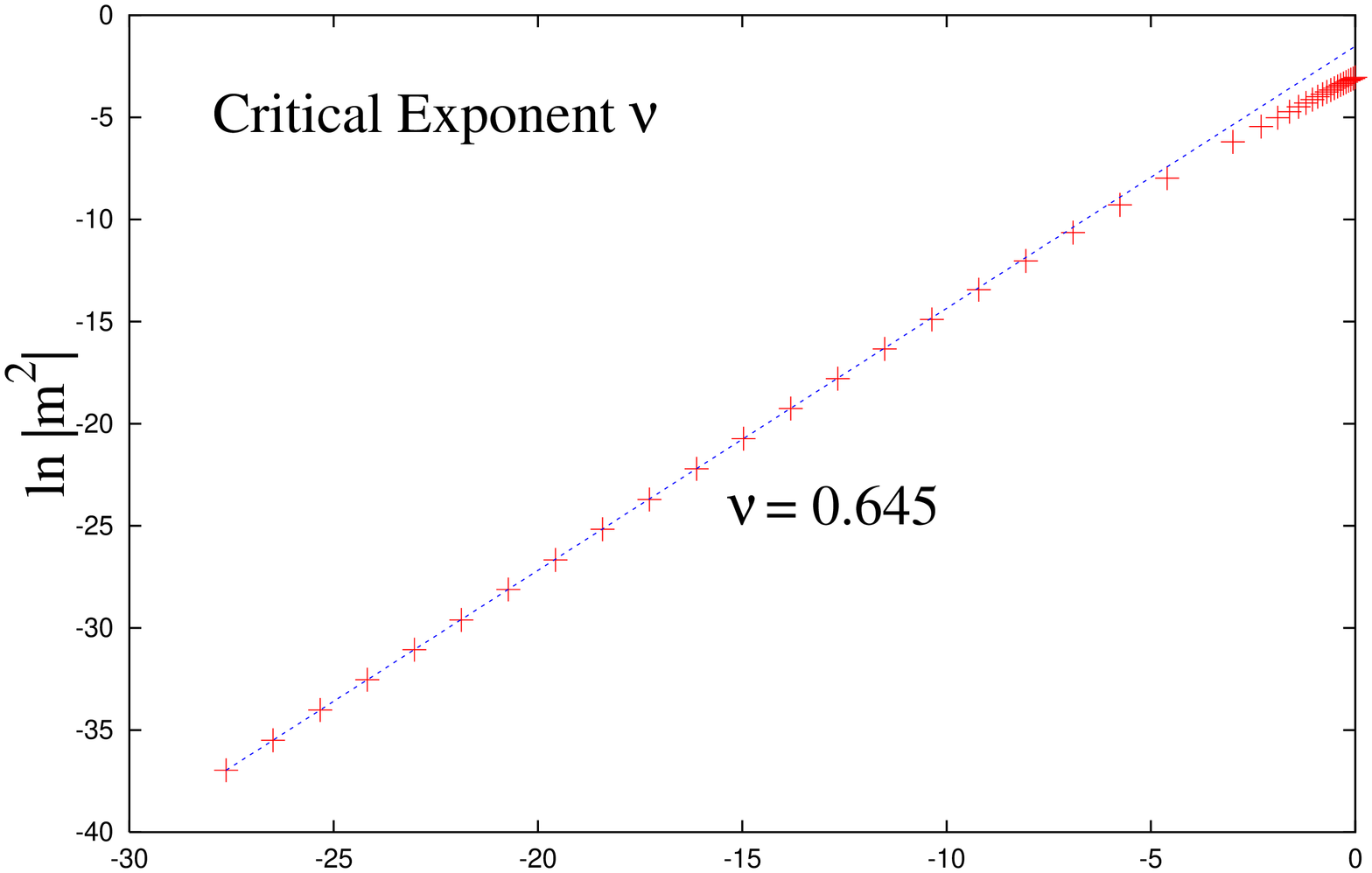 , width=2.5in , angle=-90}
\\
\vspace{-0.7cm}
\begin{center}
\quad
ln $\left| {\phi_0^2} |_{t=0} -  {\phi_0^2} |_{t=0}^{cr} \right|$
\end{center}
\parbox{12cm}
{\caption{\label{fignu}
The critical exponent $\nu$ for $N=4$.}}
\end{figure}

The critical exponent $\nu$ describes the scaling behavior in the
critical region of the renormalized mass which is the inverse of the
correlation length $\xi = (T-T_c)^{-\nu}$ and can be defined for
arbitrary $N$ by the relation
\ba
\label{nu}
m^2 &\sim & \left| {\phi_0^2} |_{t=0} 
            -  {\phi_0^2} |_{t=0}^{cr} \right|^{2 \nu} 
\ea
with the renormalized mass in the symmetric phase $m^2 = k^2 u' (0)$
for $k\to0$. Wavefunction renormalization corrections are included in
the potential. Exactly at the critical temperature i.e.~for
$(\left. {\phi_0^2} \right|_{t=0}=\left. {\phi_0^2}
\right|_{t=0}^{cr})$ the renormalized mass vanishes and hence the
correlation length diverges. For the determination of the exponent
$\nu$ we calculate $m^2$ for different values of
$\left.{\phi_0^2}\right|_{t=0}$ around
$\left. {\phi_0^2}\right|_{t=0}^{cr})$ and plot $\ln(m^2)$ as a
function of $\ln \left|\left.{\phi_0^2}\right|_{t=0}-\left.{\phi_0^2}
\right|_{t=0}^{cr}\right|$. The correlation is linear (see
Fig.~\ref{fignu}) and the slope yields the exponent $\nu$. 
\noindent
Note that the definition of the renormalized mass $m^2$ results in a positive 
value only in the symmetric phase. Therefore one has to make sure to reach the 
symmetric phase via ${\phi_0^2} (t) \to 0$ before the evolution with respect 
to $t$ is stopped.

\begin{figure}
\begin{center}
\begin{tabular}{|c|c|c|c|c|c|c|}
\hline
$\barr{c}\ \\ \ \earr$  & $\eta$  & $\delta$ & $\nu$ & $\beta$  \\
\hline\hline
N=0 $\barr{c} \mbox{RG} (1) \\ \mbox{RG} (2) \\ \mbox{lattice}  \\ \eps \\ SPS
\earr$  
& $\barr{c} 0.039 \\ 0.034 \\ 0.041(25) \\ 0.032(3) \\ 0.027(4) \earr$    
& $\barr{c} 4.77 \\ 4.80 \\ 4.77(14) \\ 4.81(2) \\ 4.84(3) \earr$  
& $\barr{c} 0.59 \\ \ \\ 0.592(3) \\ 0.588(2) \\ 0.588(2) \earr$  
& $\barr{c} 0.30 \\ \ \\ 0.308(6) \\ 0.304(2) \\ 0.302(2) \earr$\\  
\hline
N=1 $\barr{c} \mbox{RG} (1) \\ \mbox{RG} (2) \\ \mbox{lattice} \\ \eps \\ SPS 
\earr$  
& $\barr{c} 0.044 \\ 0.045 \\ 0.044(31) \\ 0.038(3) \\ 0.032(3) \earr$
& $\barr{c} 4.75 \\ 4.74 \\ 4.75(17) \\ 4.78(2) \\ 4.81(2)  \earr$
& $\barr{c} 0.64 \\ \ \\ 0.631(2) \\ 0.631(2) \\ 0.630(2) \earr$
& $\barr{c} 0.32 \\ \ \\ 0.329(9) \\ 0.327(2) \\ 0.325(2) \earr$\\
\hline
N=2 $\barr{c} \mbox{RG} (1) \\ \mbox{RG} (2) \\ \mbox{lattice} \\ \eps \\ SPS 
\earr$  
& $\barr{c} 0.043 \\ 0.042 \\ 0.038(18) \\ 0.040(3) \\ 0.033(4) \earr$
& $\barr{c} 4.75 \\ 4.76 \\ 4.78(10) \\ 4.77(2) \\ 4.81(3) \earr$   
& $\barr{c} 0.69 \\ \ \\ 0.674(2) \\ 0.671(5) \\ 0.670(2) \earr$
& $\barr{c} 0.35 \\ \ \\ 0.350(5) \\ 0.349(4) \\ 0.346(2) \earr$\\
\hline
N=3 $\barr{c} \mbox{RG} (1) \\ \mbox{RG} (2) \\ \mbox{lattice} \\ \eps \\ SPS 
\earr$  
& $\barr{c} 0.041 \\ 0.038 \\ 0.038(17) \\ 0.040(3) \\ 0.033(4) \earr$
& $\barr{c} 4.76 \\ 4.78 \\ 4.78(10) \\ 4.77(2) \\ 4.81(3) \earr$   
& $\barr{c} 0.74 \\ \ \\ 0.711(2) \\ 0.710(7) \\ 0.705(3) \earr$
& $\barr{c} 0.37 \\ \ \\ 0.369(5) \\ 0.368(4) \\ 0.365(3) \earr$\\
\hline
N=4 $\barr{c} \mbox{RG} (1) \\ \mbox{RG} (2) \\ \mbox{lattice} \\ \eps
\earr$  
& $\barr{c} 0.037 \\ 0.034 \\ 0.0254(38) \\ 0.03(1) \earr$
& $\barr{c} 4.79 \\ 4.80  \\ 4.851(22) \\ 4.82(6) \earr$  
& $\barr{c} 0.78 \\ \ \\ 0.7479(90) \\ 0.73(2) \earr$
& $\barr{c} 0.40 \\ \ \\ 0.3836(46) \\ 0.38(1) \earr$\\
\hline
N=7 $\barr{c} \mbox{RG} (1) \\ \mbox{RG} (2)  \earr$  
& $\barr{c} 0.027 \\ 0.023 \earr$
& $\barr{c} 4.84 \\ 4.87 \earr$
& $\barr{c} 0.86 \\ \ \earr$  
& $\barr{c} 0.43 \\ \ \earr$ \\
\hline
N=10 $\barr{c} \mbox{RG} (1) \\ \mbox{RG} (2)  \earr$ 
& $\barr{c} 0.021 \\ 0.017 \earr$
& $\barr{c} 4.88 \\ 4.90 \earr$
& $\barr{c} 0.91 \\ \ \earr$  
& $\barr{c} 0.45 \\ \ \earr$ \\
\hline
N=100 $\barr{c} \mbox{RG} (1) \\ \mbox{RG} (2)  \earr$
& $\barr{c} 0.0025 \\ 0.002 \earr$
& $\barr{c} 4.99 \\ 4.99 \earr$
& $\barr{c} 0.99 \\ \ \earr$  
& $\barr{c} 0.49 \\ \ \earr$ \\
\hline\hline
$\barr{c}\ \\ \  \earr$\mbox{large-$N$} & 0 & 5      & 1.0    & 0.5     \\
\hline
\end{tabular}
\end{center}
\begin{center}
\parbox{13cm}{Table 2:{\label{tab3} Critical exponents for different $N$
compared with a perturbative calculation ($SPS$), the
$\eps$--expansion ($\eps$)~\cite{zinn} and lattice
results~\cite{kana}. RG(1) and RG(2) denote our results.  See text for
details.  }}
\addtocounter{figure}{-1}
\addtocounter{table}{1}
\end{center}
\end{figure}

In statistical physics the exponent $\delta$ describes the variation
of the magnetization (here $\Phi$) in an external field $B$ in the limit
$B\to0$ at $T_c$:
\be
B  = \f{\partial V (\Phi)}{\partial \Phi} \sim \Phi^\delta\ .
\ee
This means that the rescaled potential $u ({\phi^2})$ is proportional
to ${\phi}^{\delta+1}$ for small fields. In mean-field theory where
no quantum fluctuations are taken into account the potential 
near the origin starts with a $\phi^4$-term in four-dimensions 
at the critical temperature. This behavior changes completely in
three dimensions. In order to determine the critical exponent $\delta$
we study the flow equations in the limit ${\phi^2} \to \infty$
corresponding to $k\to0$ for fixed (not rescaled) $\Phi^2$. In this
limit $u'$ also increases and we can therefore simplify the flow
equations by neglecting the threshold functions (see
Eq.~(\ref{flowpotcrit})). At the scaling solution $\partial_t u = 0$
(corresponding to $T = T_c$) we can solve the equation
analytically to find
\be
u_* ({\phi^2}) \sim {\phi}^{2d/(d-2+\eta)}\ .
\ee
yielding the well-known scaling relation between $\delta$ and $\eta$
(see Eq.~(\ref{scalingrelation}) and \cite{tetr}).

In order to extract the exponent $\delta$ numerically we calculate the 
first derivative of the potential $u$ w.r.t. the rescaled field
${\phi^2}$ and determine the slope of $\ln u'$ for large
$\ln {\phi^2}$.

Table~2 summarizes the resulting critical exponents $\eta$, $\delta$,
$\nu$ and $\beta$ for various $N$ within our RG approach (without
quoting small numerical errors). The label RG(1) denotes the numerical
results for the exponents $\eta$, $\nu$ and $\beta$ which we have
obtained from the scaling solution (cf.~e.g.~Fig.~\ref{figscaling}).

Via the scaling relations (\ref{scalingrelation}) we can, on the other
hand, directly calculate $\delta$ once knowing $\eta$. The
abbreviation RG(2) labels the calculated result for the exponent
$\eta$. We also compare these exponents with a lattice study
\cite{kana}, an $\epsilon$--expansion (denoted by $\epsilon$) and to a
Borel-resummed perturbative calculation up to three-loops at fixed
dimension three (labeled by $SPS$) \cite{zinn}.

One nicely observes the convergence of all calculated critical
exponents to the large-$N$ values which are quoted in the last line of
Table~2. It is known that in the large-$N$ limit, the local potential
approximation becomes exact and the anomalous dimension $\eta$
vanishes \cite{morri}. For the direct determination of the remaining
critical exponent $\alpha$ we would have to calculate the specific
heat at the critical temperature which was done in
\cite{scha}. We omit the quotation of this exponent in
Table~2.

\subsection{The influence of the blocking functions}
\label{infblocking}

The structure of the regularized RG-improved flow equations (but not
the universal results) depends on the $a$ $priori$ unknown cutoff
function $f_k^{(i)}$.  The index $i$ refers to the order of
derivatives $(\partial^{2i})$ in the gradient expansion of the
action. Thus blocking functions with $i>1$ regularize higher-order
contributions to the action which go beyond the uniform wavefunction
renormalization approximation, used here.

In a previous paper \cite{papp} we have studied different choices of
the cutoff function for a truncated set of flow equations (A similar
analysis was done in \cite{aoki}). In order to see this influence for
the non-truncated set of flow equations for the full potential we
derive flow equations for a similar choice of the cutoff functions as
in ref.~\cite{papp} and calculate some critical exponents.

To be consistent with the structure of the higher momentum integrals
we evaluate differential equations for each higher blocking
function. Without the explicit solution of these blocking-function
equations we can still find the corresponding set of flow equations
(without the wavefunction contribution). We list below the
corresponding flow equations for cutoff functions $i=0,1,2$ although
$f_k^{(2)}$ is not needed in our approximation scheme:\footnote{See
notation and definitions in Section~\ref{section32}.}

\ba
f_k^{(0)}\ : \qquad \partial_t V &=& -\f{k^d}{2} S_d \left[ \ln (k^2 +
m^2_\sigma) + (N-1) \ln (k^2 + m^2_\pi) \right]\\
f_k^{(1)}\ : \qquad \partial_t V &=& \f{k^d}{d} S_d \left[
\f{1}{(1+m^2_\sigma/k^2)} + \f{N-1}{(1+m^2_\pi/k^2)} \right]\\
f_k^{(2)}\ : \qquad \partial_t V &=& \f{2 k^d  S_d}{d(d+2)}  \left[
\f{1}{(1+m^2_\sigma/k^2)^2} + \f{N-1}{(1+m^2_\pi/k^2)^2} \right]
\ea

As an example we take $N=4$ and compare the dependence of the
critical exponents $\nu$ and $\beta$ on the blocking function of order $i$.  

\begin{table}[!htb!]
\begin{center}
\begin{tabular}{|l|c|c|}
\hline
$i$ & $\nu$ & $\beta$ \\
\hline
  0   & 0.853  & 0.42  \\
  1   & 0.815  & 0.405 \\
  2   & 0.814  & 0.403 \\
\hline
\end{tabular}
\end{center}
\begin{center}
\parbox[t]{7cm}{\caption{\label{tab2} The critical exponents 
$\nu$ and $\beta$ for $N=4$ with different blocking functions
$f^{(i)}_k$ (without wavefunction renormalization).}}
\end{center}
\end{table}

In Table~{\ref{tab2}} the results are listed for various cutoff
functions $f_k^{(i)}$. A small systematic decrease in the values with
the order $i$ is observed. Already for $i\ge1$ there is almost no
difference which validates the findings in \cite{papp}. This permits
the conclusion that for higher blocking functions the convergence
becomes more and more stable and a decrease towards the values found
by lattice simulations and $\eps$-expansions is observed.

\section{Summary and conclusions}
\label{conclusion}

Near a phase-transition a perturbative treatment fails as an adequate
description of the critical behavior and nonperturbative methods such
as the RG method become necessary.  We have presented a RG study for
the finite-temperature phase transition of self-interacting scalar
theories with $O(N)$-symmetry. We have used an implementation of the
Wilsonian approach to field theories in thermal equilibrium which
provides a direct link of the $O(N)$ universal behavior near the
critical temperature with the physics at zero temperature. It offers a
nonperturbative way of studying the mechanisms of dimensional
reduction.

Based on an application of Schwinger's proper-time (heat kernel)
regularization within a one-loop expansion of the effective Euclidean
action we have introduced various cutoff functions by comparing the
proper-time regularization with a sharp-momentum regularization rendering
the resulting flow equations infrared finite. This specific choice of
cutoff functions has the additional advantage that all resulting flow
equations can be expressed in a transparent and analytical form.

We have generalized previous studies \cite{scha},\cite{papp} to a
general $O(N)$-symmetry by going beyond the local potential
approximation in a derivative expansion. We have solved the highly
coupled set of flow equations numerically on a grid without an
specific ansatz or truncation for the potential and its derivatives
during the evolution. In the infrared a convex potential is obtained
for all temperatures, consistent with the concept of the effective
action approach which is based on a Legendre transformation.

At the critical temperature all dimensionful
quantities can be rescaled which simplifies the numerics. The
dimensional reduction phenomenon which is embedded in a very
transparent way in our threshold functions could be confirmed. 
We have discussed the
critical behavior of the $O(N)$ model in detail, by determining fixed
points and beta functions for various $N$. We have calculated independently
four critical exponents verifying the scaling relations among them.

One important ingredient in our approach is the choice of the blocking
function which is $a$ $priori$ unknown and could influence the
universal quantities via the convergence properties of the 
flow equations. In order to control this influence we have
recalculated two critical exponents ($\nu$ and $\beta$) for $N=4$ for
three different choices of the blocking functions omitting the
wavefunction renormalization. No strong dependence has been observed. 
Within numerical errors we find stable exponents except for the
lowest blocking function $f_k^{(0)}$ due to a poor convergence. For
higher blocking functions the convergence becomes more and more stable
in agreement with \cite{papp}.

Altogether the presented RG approach has proven to be a powerful
nonperturbative tool in the study of finite-temperature field
theory. Due to the omission of any temperature dependence of the
initial conditions in the ultraviolet its reliability should be best
for low temperature. For $N=4$ our results indeed agree perfectly with
chiral perturbation theory. In contrast to chiral perturbation theory
we can, however, extend the temperature range to the critical
temperature where the phase transition takes place. This encourages
the application of this approach to other theories involving
e.g. gauge-fields.

\section*{Acknowledgments}
One of the authors (B.J.S.) would like to thank Bastian Bergerhoff,
Michael Buballa, Micaela Oertel and Gabor Papp for useful
discussions. This work was supported in part by the GSI and DFG.

\end{fmffile}

\begin{appendix}
\section{Appendix}
\label{apptech}  
\setcounter{section}{0}
\setcounter{equation}{0}
\setcounter{table}{0}
\renewcommand{\thesection}{A.\arabic{section}}
\renewcommand{\theequation}{A.\arabic{equation}}
\renewcommand{\thetable}{A.\arabic{table}}

In order to improve the readability of Sec.~\ref{section3} we collect
in this appendix some technical details concerning the derivation of
the wavefunction renormalization flow equation (\ref{wavefunction}).

The expansion of the effective $O(N)$-invariant Lagrangian to
two\--de\-riva\-ti\-ve order (see Eq.~(\ref{lagordertwo})) requires
the calculation of a trace over ($N \times N$)-matrix valued
terms stemming from different powers of the second derivative of
the potential $V''_{ij}$ with $i,j = 1, \ldots, N$.

This can basically be accomplished by introducing projectors
\begin{eqnarray*}
X_1 = \f{\Phi_i \Phi_j}{\vec{\Phi}^2} &\mbox{and}& 
X_2 = \delta_{ij} - \f{\Phi_i \Phi_j}{\vec{\Phi}^2}\ ,
\end{eqnarray*}
which satisfy the usual projection operator property $X_a X_b =
\delta_{ab} X_b$ for $a, b = 1,2$ and $tr X_1 = 1$, $tr X_2 = N-1$.

Using the abbreviations $m^2_\sigma \equiv \lambda (3 \vec{\Phi}^2 -
\Phi_0^2 )$ and $m^2_\pi \equiv \lambda (\vec{\Phi}^2 - \Phi_0^2 )$
(cf.~Eq.~(\ref{massabb})) the $N \times N$ matrix $V_{ij}''$
can be rewritten by
\begin{eqnarray*}
V_{ij}''= \lambda \left( \vec{\Phi}^2 -\Phi_0^2 \right) \delta_{ij} + 2
\lambda \Phi_i \Phi_j \  = m_\sigma^2 X_1 + m_\pi^2 X_2\ .
\end{eqnarray*}
Using the projector properties different powers of $V_{ij}''$ can be readily 
evaluated. One finds, for example,
\ba
\label{long}
&& Tr \left[ \sum_{n=2}^{\infty} \frac{ (-1)^{n-1} \tau^n}{n!} \sum_{k=0}^{n-1}
\left( V'' \right)^k \partial^2_\mu \left( V'' \right)^{n-1-k} \right] \nonumber \\
&& \quad = -\left( \frac 12 \tau^2 \partial^2_\mu m_{\sigma}^2 - \frac 13 \tau^3 
\partial_{\mu} m_{\sigma}^2 \partial_{\mu} m_{\sigma}^2 \right) 
e^{- \tau m_{\sigma}^2 } 
Tr \left[ X_1 \right] \\
&& \quad \quad -\left( \frac 12 \tau^2 \partial^2_\mu m_{\pi}^2 - \frac 13 \tau^3 
\partial_{\mu} m_{\pi}^2 \partial_{\mu} m_{\pi}^2 \right) e^{- \tau m_{\pi}^2 } 
Tr \left[ X_2 \right] \nonumber \\
&& \quad \quad -\left( \frac{2}{m_{\sigma}^2 - m_{\pi}^2 } \left( e^{- 
\tau m_{\pi}^2 } - e^{- \tau m_{\sigma}^2 } \right) -\tau \left( e^{- \tau m_{\sigma}^2 }
 + e^{- \tau m_{\pi}^2 } \right) \right) Tr \left[ X_1 \partial^2_\mu X_1
\right]\nonumber
\ea
where on the l.h.s.~of Eq.~(\ref{long}) all finite and infinite sums
are analytically calculable (cf.~\cite{oles}) and the remaining traces
are given by a straightforward calculation as
\begin{eqnarray*}
Tr \left[ X_a \partial_\mu X_b \right] &=& 0 \qquad \mbox{with} \qquad 
a,b,c = 1,2\ ,\\
Tr \left[ X_a \partial^2_\mu X_b \right] &=& 2 (-1)^{\delta_{ab}+1} 
\left(  \frac{\left( \vec{\Phi} \partial_\mu \vec{\Phi} \right)^2}{
\left(\vec{\Phi}^2 \right)^2} - \frac{ (\partial_\mu \vec{\Phi} )^2}
{\vec{\Phi}^2 } \right)\ , \\
Tr \left[ X_a \partial_\mu X_b \partial_\mu X_c \right] &=& (-1)^{\delta_{bc}}
\left( \frac{\left( \vec{\Phi} \partial_\mu \vec{\Phi} \right)^2}{
\left(\vec{\Phi}^2 \right)^2} - \frac{ (\partial_\mu \vec{\Phi})^2}
{\vec{\Phi}^2 } \right)\ .
\end{eqnarray*}
Evaluating the other contributions to the effective action in the same
manner yields Eq.~(\ref{gamma2}).

\end{appendix}


\end{document}